\documentclass[aps,twocolumn]{revtex4-1}
\pdfoutput=1

\usepackage{amsmath}
\usepackage{amssymb}

\usepackage{graphicx}

\usepackage{color}

\date{}

\usepackage{bussproofs}

\usepackage[all]{xy}
\usepackage{hyperref}
\hypersetup{colorlinks=true,
linkcolor=[rgb]{.67 .27 .27},
citecolor=[rgb]{.09 .29 .54},
urlcolor=[rgb]{.09 .29 .54}}
\usepackage[doipre={doi:}]{uri}

\usepackage[table]{xcolor}
\usepackage{rotating}
\usepackage{listings}
\usepackage{setspace}
\definecolor{Code}{rgb}{0,0,0}
\definecolor{Decorators}{rgb}{0.5,0.5,0.5}
\definecolor{Numbers}{rgb}{0.5,0,0}
\definecolor{MatchingBrackets}{rgb}{0.25,0.5,0.5}
\definecolor{Keywords}{rgb}{.67, .27, .27}
\definecolor{self}{rgb}{0,0,0}
\definecolor{Strings}{rgb}{.09,.29,.54}
\definecolor{Comments}{rgb}{0,0.5,0}
\definecolor{Backquotes}{rgb}{0,0,0}
\definecolor{Classname}{rgb}{0,0,0}
\definecolor{FunctionName}{rgb}{0,0,0}
\definecolor{Operators}{rgb}{0,0,0}
\definecolor{Background}{rgb}{0.98,0.98,0.98}

\lstnewenvironment{python}[1][]{
\lstset{
numbers=left,
numberstyle=\footnotesize,
numbersep=1em,
xleftmargin=1em,
framextopmargin=2em,
framexbottommargin=2em,
showspaces=false,
showtabs=false,
showstringspaces=false,
frame=l,
tabsize=4,
basicstyle=\ttfamily\small\setstretch{1},
backgroundcolor=\color{Background},
language=Python,
commentstyle=\color{Comments}\slshape,
stringstyle=\color{Strings},
morecomment=[s][\color{Strings}]{"""}{"""},
morecomment=[s][\color{Strings}]{'''}{'''},
morekeywords={import,from,class,def,for,while,if,is,in,elif,else,not,and,or,print,break,continue,return,True,False,None,access,as,,del,except,exec,finally,global,import,lambda,pass,print,raise,try,assert},
keywordstyle={\color{Keywords}\bfseries},
morekeywords={[2]@invariant},
keywordstyle={[2]\color{Decorators}\slshape},
emph={self},
emphstyle={\color{self}\slshape},
}}{}

\usepackage{placeins}
\usepackage{longtable}
\usepackage{dot2texi}
\usepackage{tikz}
\usetikzlibrary{automata,shapes,arrows}
\usepackage[colorinlistoftodos]{todonotes}
\usepackage{amsthm}
\usepackage{bibunits}

\definecolor{DeepRed}{rgb}{.82,.14,.16}
\definecolor{DeepBlue}{rgb}{0,0.36,0.62}

\makeatletter
\def\@biblabel#1{}

\makeatother

\theoremstyle{definition}

\theoremstyle{remark}

\renewcommand*{\sectionautorefname}{Sec.}

\newcommand{\beginsupplement}{        \setcounter{table}{0}
        \renewcommand{\thetable}{S\arabic{table}}        \setcounter{figure}{0}
        \renewcommand{\thefigure}{S\arabic{figure}}        \setcounter{equation}{0}
        \renewcommand{\theequation}{S\arabic{equation}}		\setcounter{section}{0}
		\renewcommand{\thesection}{S\arabic{section}}
        \renewcommand*{\sectionautorefname}{Sec.}
        \setcounter{page}{1}
     }

\renewcommand{\thesection}{\arabic{section}}

\def\hier{\mathbf{Hier}}
\def\Vertex{\mathrm{Vertex}}
\def\adj{\mathrm{adj}}

\def\connectivity{d}
\def\reffigexamplesystemmodules{\ref{fig:modsccsym}A}
\def\reffigscc{\ref{fig:modsccsym}B}
\def\reffighiertransformations{\ref{fig:modsccsym}C}
\def\reffigrobustconnect{\ref{fig:combined}A}
\def\reffigrobusthierarchy{\ref{fig:combined}B}
\def\reffigconnectcycle3D3x3{\ref{fig:combined}C}
\def\reffigconnectdist3D3x3{\ref{fig:combined}D}
\def\refsupp{Supporting Information}

\usepackage{flushend}

\begin{document}

\let\ref\autoref

\pagenumbering{arabic}

\title{\bf Hierarchical Network Structure Promotes Dynamical Robustness}

\author{Cameron Smith$^{1}$,
Raymond S. Puzio$^{1}$,
Aviv Bergman$^{1,2,3,4,*}$}

\affiliation{$^1$Department of Systems and Computational Biology,\\
$^2$Dominick P. Purpura Department of Neuroscience,\\
$^3$Department of Pathology, Albert Einstein College of Medicine,\\
1301 Morris Park Ave, Bronx, NY 10461, USA\\
$^4$Santa Fe Institute, 1399 Hyde Park Road, Santa Fe, NM 87501, USA
\\
$*$to whom correspondence should be addressed: aviv@einstein.yu.edu}

\date{\today}

\begin{abstract}
The relationship between network topology and system dynamics has significant implications for unifying our understanding of the interplay among metabolic, gene-regulatory, and ecosystem network architecures. Here we analyze the stability and robustness of a large class of dynamics on such networks. We determine the probability distribution of robustness as a function of network topology and show that robustness is classified by the number of links between modules of the network. We also demonstrate that permutation of these modules is a fundamental symmetry of dynamical robustness. Analysis of these findings leads to the conclusion that the most robust systems have the most hierarchical structure. This relationship provides a means by which evolutionary selection for a purely dynamical phenomenon may shape network architectures across scales of the biological hierarchy.
\end{abstract}

\maketitle

\setcounter{secnumdepth}{4}

\section{Introduction}

\begin{figure*}[!ht]
\centering
\noindent\includegraphics[width=1.0\textwidth]{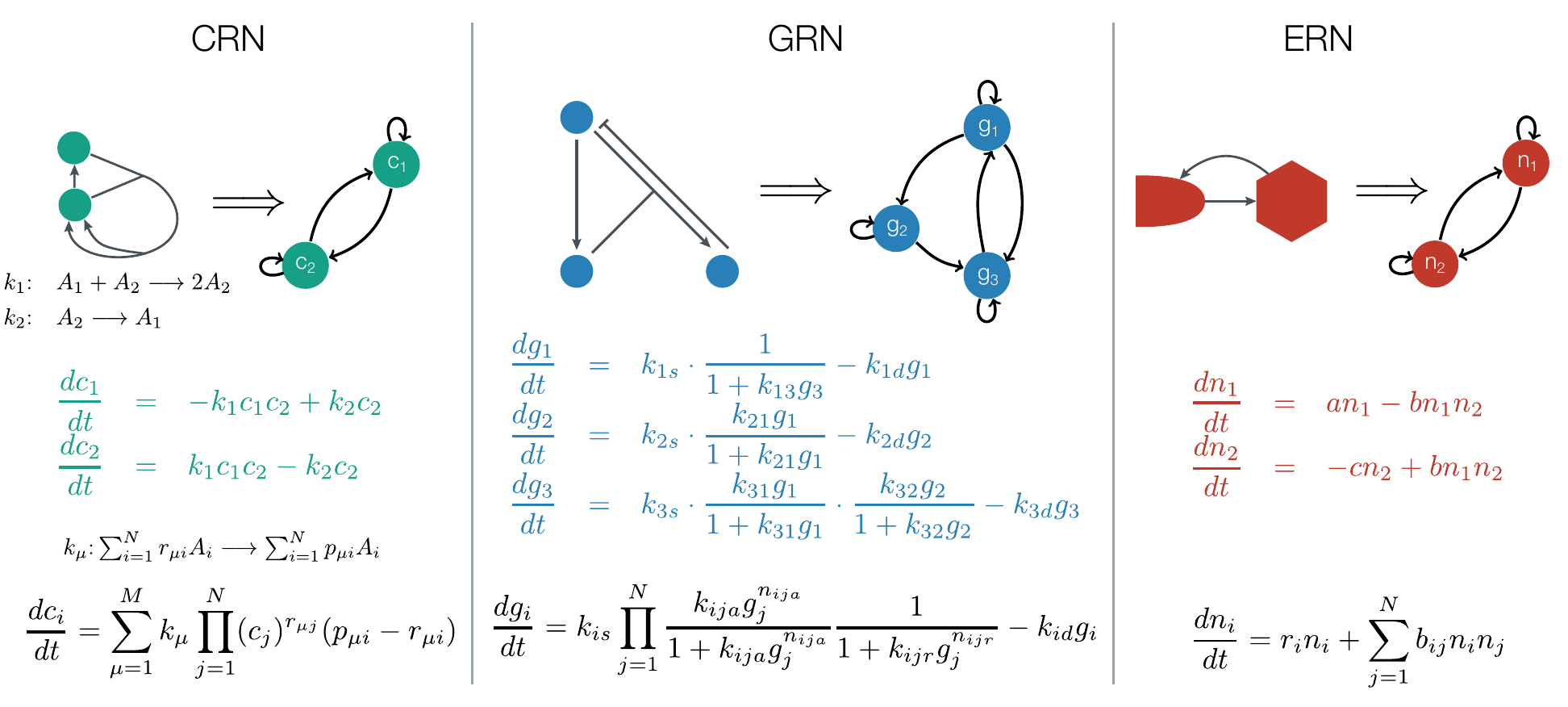}
\caption{{\bf Dynamical models in systems biology.} The top row represents a chemical reaction network (CRN) \cite{Shinar2010}, a gene regulatory network (GRN) \cite{Karlebach2008}, and an ecological regulatory network (ERN) \cite{Rohr2014} in terms of the graphical methods specific to each field mapped into the interaction graph, which provides a unified representation for networks across these fields. The second row represents a particular example of a system of differential equations that are used to model a biological network within each of the domains of application considered here. The third row shows the general form of a system of differential equations that can be used to model any network architecture within each domain.}
\label{fig:biomodelexamples}
\end{figure*}

The traditional approach taken in the study of chemical reaction, gene-regulatory, population, and ecosystem networks is to derive a system of differential equations to model a particular biological network, attempt to fit that model to data and adjust the modeling assumptions along with parameter values until a good fit is achieved \cite{Meyer2014}. Over evolutionary timescales, one expects to observe changes in the model of best fit. All of these models utilize essentially equivalent mathematical structures, instances of which are sampled in the evolutionary process (\ref{fig:biomodelexamples}, \cite{RossCr2003,Palsson2011a,Sauro2012}). Developing unified mathematical descriptions of each of these that can be embedded into models of the evolutionary process is one of the paramount goals of systems biology.

Recent work has demonstrated that as a result of the existence of largely insensitive directions within the parameter space for such models, the approach outlined above often allows for a large variety of models to fit equivalent data \cite{Machta2013,Hines2014,Prabakaran2014,Tonsing2014}. In addition, there is often uncertainty about the very structure of such networks. Since the evolutionary process results in modifications to the underlying model this fact may be used to characterize evolutionarily effective versus neutral spaces. In this context, it is crucial to gain insight into what dynamical phenomena are possible to observe within a given class of dynamical systems. This is necessary to understand in order to determine whether or not a given dynamical phenomenon should be regarded as unique or generic in the development and investigation of models applied to particular systems \cite{Gunawardena2013,Gunawardena2014}. This can be achieved using a method common in statistical physics involving the consideration of an ensemble of systems that, in comparison to one another, appear to have components that are randomly interlinked.

Investigating generic properties of a large class of dynamical systems was the approach taken by May in models of ecosystem dynamics \cite{Gardner1970,May1972}. The class of dynamical systems studied by May is, however, not restricted to ecosystem dynamics and encompasses, among others, the dynamics of all of the networks represented in \ref{fig:biomodelexamples}. May conjectured on the basis of results from random matrix theory what eventually came to be referred to as the May-Wigner stability theorem \cite{Cohen1984,May1972a,Radius2014,Majumdar2014}, which implies a relationship between a topological property, system connectivity, and a dynamical property, stability.

Here we determine the relationship between network hierarchy, a topological property, and \emph{robustness}, a dynamical property. Robustness is of interest in biological systems at all scales, and has been previously studied in the context of biochemical networks \cite{Alon1999,Shinar2010}, gene-regulatory networks \cite{VanNimwegen1999,Siegal2002,Draghi2010,Wagner2013}, and ecological networks \cite{Rohr2014}. Over physiological timescales, the robustness of a particular network state may be evaluated by determining its linear stability \cite{Davis1962}. Network states that are linearly stable, are robust to perturbations in the states of their components. For example, in the case of gene-regulatory networks, a state vector containing protein molecule counts or concentrations of each gene that is stable with respect to the dynamics of the network will exponentially suppress any relatively small modifications to the state of a gene and return to the initial stable state.

\begin{figure*}[!htbp]
\centering
\noindent\includegraphics[width=1.0\textwidth]{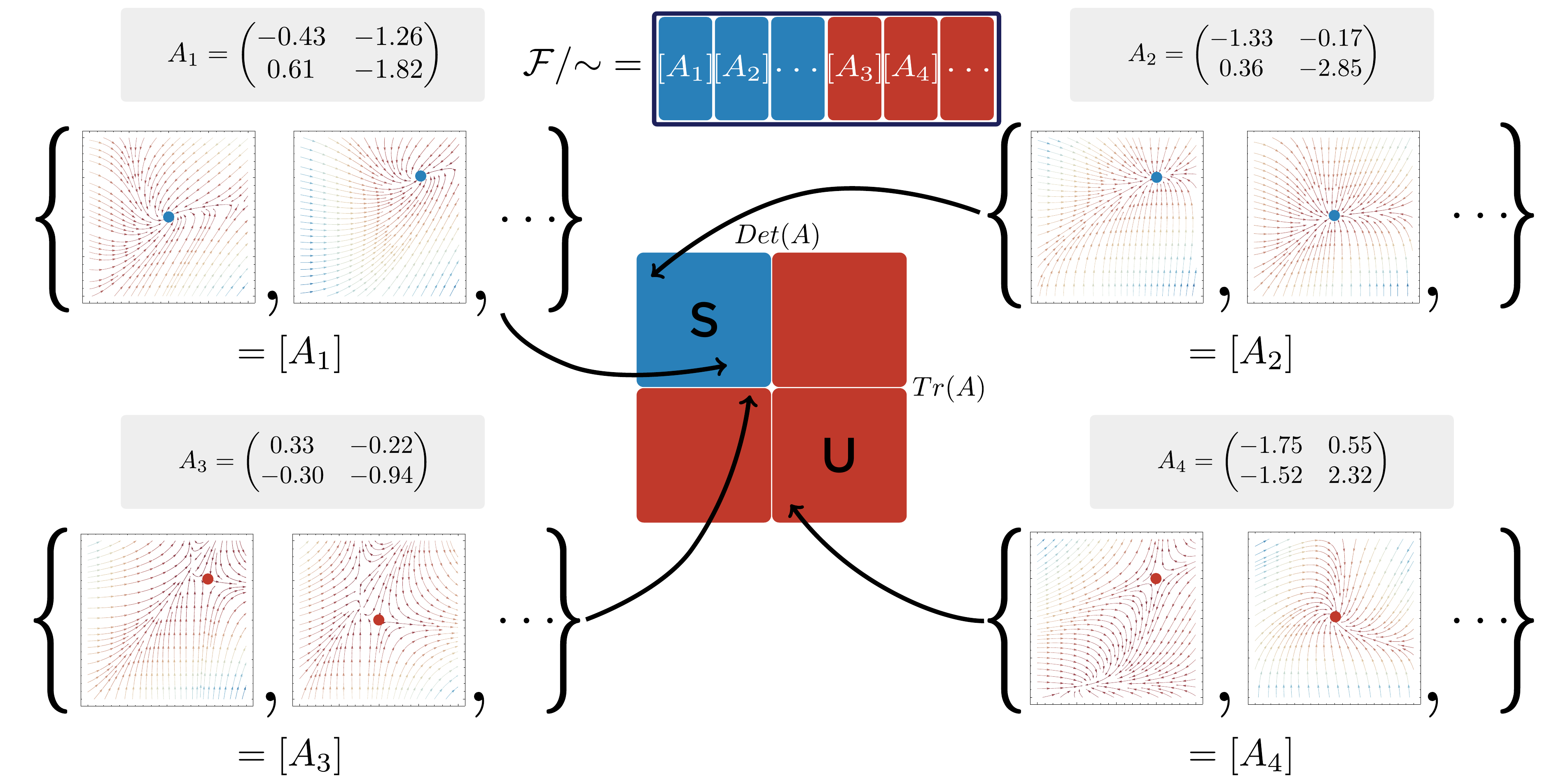}
\caption{{\bf Stability and equivalence classes of fixed points.} Linearization around a steady-state of a model of a biological network allows for the assessment of the stability of that state. For two-dimensional systems, assessment of stability in terms of the trace and determinant of the Jacobian matrix associated to a particular steady-state geometrically partitions the plane into a stable region (S, blue) and an unstable region (U, red). For larger systems in higher dimensions this geometry is too complicated to visualize, but it is nevertheless well-defined and can be investigated algebraically and numerically. (top center) Jacobian matrices such as $A_1$ partition the space of fixed points into equivalence classes $[A_1]$, each of which is either stable or unstable.}
\label{fig:robustnessconcept}
\end{figure*}

Rather than evaluating system stability for a particular model of a biological network over physiological timescales, we are interested in evaluating robustness over evolutionary timescales where the form of the most accurate underlying model is itself subject to change. Over evolutionary timescales it is expected for there to be fluctuations, not only in the state, but in any aspect of the model specifying the dynamics of the network itself (i.e. parameters, structure of the rate functions, etc.), and these changes may occur at any level of the hierarchy including metabolic, gene-regulatory, population, and ecosystem. It is therefore also expected that network architectures managing to persist over such timescales may be required to do so with modifications to the location of their stable states and even to the geometry of their state spaces. However, what must remain invariant on such evolutionary timescales is the higher-level property that the system possess a relatively high overall probability of remaining in a stable state upon modifications to the underlying dynamical process subject to environmental constraints. This is necessary in order for networks of lower-level components to exist, regardless of what state they exist in, long enough to serve as the substrate out of which networks of relatively higher-level components are constructed \cite{Simon2002}. What is important over at least moderate evolutionary timescales is then the conditional probability, $R$, given a system is in a stable state, that upon modifications to its structure in the context of environmental fluctuations, it remains stable, regardless of where the stable state is located within the state space \ref{fig:robustnessconcept}. For the purpose of this investigation then, we quantify dynamical robustness in this way.

We demonstrate that systems exhibiting maximal robustness with respect to this definition have the most hierarchical network topology and explain why this results from an invariance of robustness to particular kinds of transformations of the network topology. For the purpose of formulating this result, the maximally hierarchical network is considered to be the graph associated to the total ordering (\refsupp{} and \cite{Cormen2009}). An example of this for three system components is shown in \reffigscc{} top. We use a measure of hierarchy based upon the edit distance from this maximally hierarchical network \cite{Axenovich2011}. Our results hold for networks of arbitrary size and are independent of the probability distribution from which the strengths of interaction are sampled.

\section{Dynamical Systems on Biological Networks}

In the general case of a dynamical system with $n$ components, where the components may be concentrations of chemical species, genes, or biological species, we have an $n$-dimensional vector of state variables or observables $(x_1(t), \ldots x_n(t)) = \vec{x}(t)
$
whose components are solutions to the arbitrary first order system
\begin{equation}\label{eq:eom}
\frac{dx_i(t)}{dt} = F_i(\vec{x}(t), \vec{p}), \; (i=1,\ldots,n)
\end{equation}
where $F=\{F_1,\ldots,F_i,\ldots,F_n\}$ represent, potentially nonlinear, functions of the given vector of state variables and $\vec{p} \in \mathbb{R}^s$ is the vector of $s$ parameters of the $F_i$. These parameters typically represent reaction rates or interaction strengths in chemical, gene-regulatory and ecological networks. For example, in the Lotka-Volterra model in \ref{fig:biomodelexamples}, $\vec{x} = (n_1, \ldots, n_N)$, $\vec{p}=(r_1,\ldots,r_N,b_{11},\ldots,b_{NN})$, and $F_i = r_i n_i + \sum_{j=1}^{N} b_{ij} n_i n_j$. The set of all dynamical systems, $\mathcal{D}$, for a given number of state variables $n$ and a given number of parameters $s$, is then
\begin{equation}\label{eq:setdynsys}
\mathcal{D} = \{ (F(-,\vec{p}),\vec{p}) | F \colon \mathbb{R}^n \times \mathbb{R}^s \rightarrow \mathbb{R}^n, \vec{p} \in \mathbb{R}^s \}.
\end{equation}

Fixed points are the simplest class of solutions to the dynamical system characterizing its long-term behavior. If $\vec x$ is a fixed point (i.e. $F_i(\vec{x})=0$ for all $i$), we
may proceed to ask whether it is dynamically stable.
Intuitively, dynamic stability means that, if one chooses the initial
conditions sufficiently close to the fixed point, the solution will
stay nearby.  Physically, this is important because,
if a fixed point ${\vec x}^0$ is unstable, we have zero probability of
observing the solution ${\vec x}(t) = {\vec x}^0$ in the absence of coupling to another system. The Lotka-Volterra model has two fixed points: the trivial one of all zero species $n_i=0$ and the other given implicitly by $r_i + \sum_{j=1}^{N} b_{ij} n_j = 0$. The set of all such fixed points, $\mathcal{F}$, is then
\begin{equation}\label{eq:setfixedpoints}
\mathcal{F} = \{ (d,\vec{x}^0) \; | \; d=(F(-,\vec{p}),\vec{p}) \in \mathcal{D}, \vec{x}^0 \in \mathbb{R}^n,F(\vec{x}^0,\vec{p})=0 \}.
\end{equation}

\subsection{Stability analysis of biological networks}
To determine stability, we use the Taylor series expansion of the equations of motion \ref{eq:eom} about the fixed point $\vec{x}^0$ where $\vec{y} = \vec{x} - \vec{x}^0$ by
\begin{equation}\label{eq:taylorseries}
\begin{aligned}
\frac{dx_i(t)}{dt} \approx & F_i(\vec{x}^0)
+ \sum_{j=1}^{N} \left. \frac{\partial F_i}{\partial x_j} \right|_{\vec{x} = \vec{x}^0} y_j\\
& + \frac{1}{2}\sum_{j,k=1}^{N} \left. \frac{\partial^2 F_i}{\partial x_j \partial x_k} \right|_{\vec{x} = \vec{x}^0} y_j y_k + \cdots
\end{aligned}
\end{equation}
The zeroth order term vanishes since $F_i(\vec{x}^0)=0$ by definition and thus neglecting terms higher than first order from \ref{eq:taylorseries} results in
\begin{equation}\label{eq:lineardynsys}
\frac{d\vec{y}(t)}{dt} = A \vec{y}(t),
\end{equation}
where the $n \times n$-matrix $A$ has components
$$
a_{ij} = \left. \frac{\partial F_i}{\partial x_j} \right|_{\vec{x} = \vec{x}^0}.
$$

The system defined by $F_i$, $\vec{p}$, and $\vec{x}^0$ is dynamically stable if the eigenvalues of $A$ all have real parts less than zero and $A$ is then referred to as a stable matrix. The spectral abscissa of the matrix $A$ is defined as
$$
\eta(A) = \max_i \{\Re(\lambda_i)\}
$$
where $\lambda_i$ are the eigenvalues of $A$. The system defined by $F_i$ and $\vec{x}^0$ is dynamically stable if the spectral abscissa of $A$ is less than zero, equivalently, $\eta(A) < 0$. This is because the general solution to \ref{eq:lineardynsys} is
$$
y_i(t) = \sum_j b_{ij} e^{\lambda_j t}, \; (i=1,\ldots,n)
$$
for some matrix $B=(b_{ij})$ and thus all $\vec{y} = \vec{x} - \vec{x}^0$ decay to zero when all $\lambda_i < 0$.

This criterion can be checked equivalently in terms of conditions on the coefficients of the characteristic polynomials $\chi(A)$ associated to the systems described by matrices $A$. In the $2$-dimensional case, $\chi(A) = \lambda^2 + Tr(A)\lambda+Det(A)$ has solutions $\lambda$ with negative real parts if $Tr(A)<0$ and $Det(A)>0$, which we make use of in examples. Generalized conditions for higher dimensions are available in \cite{Gantmacher1959}. As an example of a Jacobian matrix, the Lotka-Volterra model has
 \begin{equation}\label{eq:lotkavolterrajacobian}
   a_{ij} = \left\{
     \begin{array}{lr}
       r_i + b_{ii} n_i + \sum_{k=1}^{N} b_{ik} n_{k}, & i \neq j\\
       b_{ii} n_i & i=j
     \end{array}.
   \right.
\end{equation}

Evaluation of the stability criterion occurs on a space of two states inducing a mapping from matrices to binary values $\mathcal{S} \colon \mathbb{R}^{n \times n} \rightarrow \{ 1, 0 \}$ given by
 \begin{equation}\label{eq:stabeval}
   \mathcal{S}(A) = \left\{
     \begin{array}{lr}
       1, & \eta (A) < 0\\
       0, & \eta (A) \geq 0
     \end{array},
   \right.
\end{equation}
where $1$ stands for $S$ or stable and $0$ stands for $U$ or unstable. The stability criterion defines an equivalence relation on the set of all Jacobian matrices $A \in \mathbb{R}^{n \times n}$ deriving from fixed points on $n$ variables that simply splits the set into two classes $S=\{ A \, | \, A \hbox{ is stable}  \}$ and $U~=~\{ A \, | \, A \hbox{ is unstable} \}$.

\subsection{Equivalence classes of systems associated to Jacobian Matrices}
Jacobian matrices define an equivalence relation, $\sim$, on fixed points given by $(F,\vec{p},\vec{x}^0) \sim (F',\vec{p}\,',\vec{x}^{0'}) $ if and only if
\begin{equation}\label{eq:jaceqrel}
\left. \frac{\partial F(\vec{x},\vec{p})}{\partial \vec{x}} \right|_{\vec{x} = \vec{x}^0} =
\left. \frac{\partial F'(\vec{x},\vec{p}\,')}{\partial \vec{x}} \right|_{\vec{x} = \vec{x}^{0'}}.
\end{equation}
This relation then partitions the set of all fixed points into equivalence classes $\mathcal{F}/{\sim}$, where the class $[A]$ associated to Jacobian matrix $A \in \mathbb{R}^{n \times n}$ is
\begin{equation}\label{eq:jaceqs}
[A] = \left\{ (F,\vec{p},\vec{x}^0) \; | \; \left. \frac{\partial F(\vec{x},\vec{p})}{\partial \vec{x}} \right|_{\vec{x} = \vec{x}^0} = A \right\}.
\end{equation}
An example with two members of $\mathcal{F}$ from each of four different equivalence classes $[A_1]$, $[A_2]$, $[A_3]$, and $[A_4]$ of $\mathcal{F}/{\sim}$ is shown in \ref{fig:robustnessconcept}.

\subsection{Interaction graphs encoding network architecture}
The interactions among variables in a dynamical model for any network can be represented in terms of a global interaction graph (\ref{fig:biomodelexamples} top row).
For a general system $X \in \mathcal{D}$ the directed graph $G_X$ that describes the manner in which each of the variables depends upon one another is given by the adjacency matrix $\adj(G_X)$ where $\adj(G_X)_{ij}$ is $1$ if $F_i \hbox{ depends on } x_j$ and $0$ if $F_i \hbox{ does not depend on } x_j$. These two conditions on global system interactions are expressed respectively for all $\vec{x}$ in terms of elements of the Jacobian matrix as $\frac{\partial F_i}{\partial x_j}(\vec{x}) \neq 0$ and $\frac{\partial F_i}{\partial x_j}(\vec{x}) = 0$.
For large systems, since any given component is only likely to interact with a relatively small proportion of the other components, these matrices may be sparse.
We can also associate a local interaction graph $G_A$ given by an adjacency matrix $\adj(G_A)$ to each dynamical system having Jacobian matrix $A$ at some fixed point $\vec{x}^0$ where
 \begin{equation}\label{eq:lininterdepadj}
   \adj(G_A)_{ij} = \left\{
     \begin{array}{lr}
       1, & a_{ij} \neq 0\\
       0, & a_{ij} = 0
     \end{array}.
   \right.
\end{equation}
In general, the graph $G_A$ is a subgraph of $G_X$, however, $G_A$ is almost always equivalent to $G_X$ (\refsupp{}). We define the connectivity to be equal to the number of edges in $G_A$, which is equivalent to summing up the number of non-zero entries of $\adj(G_A)$.
These interaction graphs can be viewed as deriving from the combination of system components that accept a given pattern of inputs and produce a given pattern of outputs (\reffigexamplesystemmodules).

\begin{figure*}[!htbp]
\centering
\noindent\includegraphics[width=1.0\textwidth]{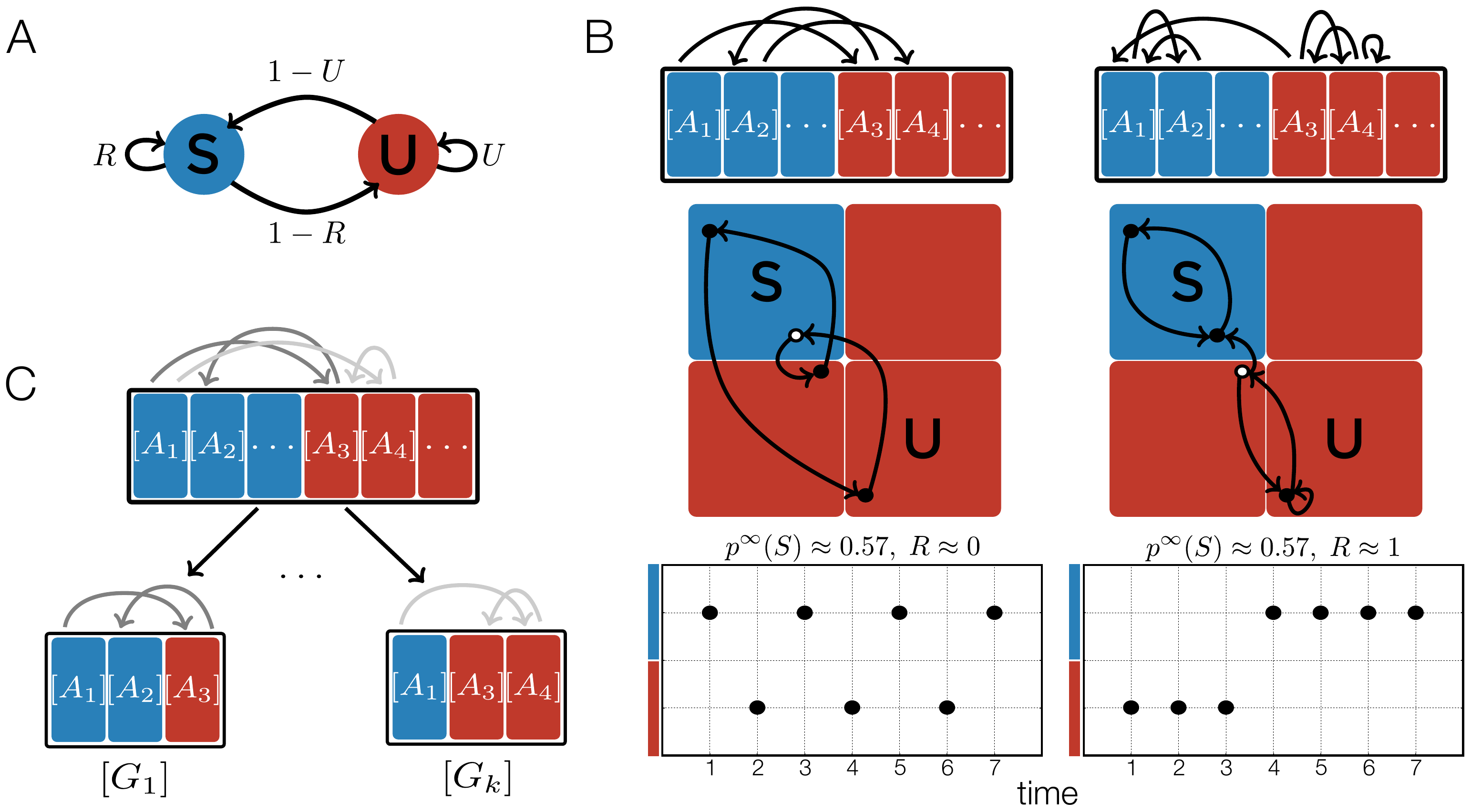}
\caption{{\bf Assessment of dynamical robustness.} (A) Any stochastic process induced on the stable and unstable regions of the space of fixed points can be represented as a state transition diagram. The conditional probability of remaining in the stable region in the context of such modifications, given a previously existing stable state, is then provided by estimating $R$. (B) The stochastic process that results from modifications to a given dynamical model over evolutionary timescales induces a stochastic process on the equivalence classes of fixed points $\mathcal{F}/{\sim}$ represented by the boxes in the top row. This in turn induces a process on the stable and unstable regions, which can be used to estimate robustness. We note that $R$ is distinct from the stationary probability of being within the stable region $P^{\infty}(S) = \frac{1-U}{2-R-U}$. (C) Each network architecture encoded by a directed graph, $\{G_1, \ldots, G_k \}$, that indicates the manner in which the variables of a biological network depend upon one another selects a particular subset of equivalence classes from $\mathcal{F}/{\sim}$ (see \ref{eq:jacgrapheqs}). Robustness can therefore be estimated independently for each network architecture.}
\label{fig:robustnessprocess}
\end{figure*}

Each distinct directed graph $G$, of which there are $k=2^{n(n-1)}$ that could be associated to the interactions in a model defined on $n$ variables, selects a subset of $\mathcal{F} / {\sim}$ (see~\ref{fig:robustnessprocess}C)
\begin{equation}\label{eq:jacgrapheqs}
[G] = \left\{ [A] \; | \; G_A = G \right\}.
\end{equation}
The $G$-classes thereby partition the collection of fixed points, $\mathcal{F}$, over the space of dynamical systems, $\mathcal{D}$, according to the interactions among the variables of the dynamical system represented by the topology of $G$. This partition, $\mathcal{F} / G$, is a coarsening of $\mathcal{F} / {\sim}$.

\section{Evolutionary processes sampling dynamical systems}
In the course of biological evolution, the parameter values $\vec{p}$, form of the functions $F$, and environmental conditions restricting access to the basins associated to different fixed points $\vec{x}^0$ corresponding to all different types of networks considered in \ref{fig:biomodelexamples} are subject to, potentially drastic, modifications due to environmental fluctuations. The stochastic process by which these modifications occur induces one on the set of fixed points that results in the assignment of a probability $P(f^T)$ to each history of length $T$, $f^T = ( f_1,f_2,\ldots,f_T ) \in \mathcal{F}^T$ \cite{RobertM.Gray130}.
These dynamics induce a stochastic process on the equivalence classes of fixed points $\mathcal{F}/{\sim}$ indexed by Jacobian matrices given by
$$
P([A]^T) = \sum_{ \{ f^T \mid f_i^T \in [A]_i^T \} } P(f^T)
$$
for each history of length $T$, $[A]^T = ( [A_1], [A_2], \ldots, [A_T] ) \in (\mathcal{F}/{\sim})^T$
as visualized in \ref{fig:robustnessprocess}B. These changes can alter the stability of a given system, thereby inducing an even more coarse-grained stochastic process on the stable regions of the space of fixed points given by
$$
P(s^T) = \sum_{ \{ A^T \mid \mathcal{S}(A_i^T) = s_i^T \} } P(A^T)
$$
for each history of length $T$, $s^T = ( s_1, s_2, \ldots, s_T ) \in \{0,1\}^T$
as visualized in \ref{fig:robustnessprocess}A and B. In order to model this we consider a process whereby perturbations applied to a given dynamical system and fixed point associated to a stable Jacobian matrix $A$ lead to another Jacobian matrix $A'$.  This corresponds to an ensemble of fixed points of dynamical systems where each model in the ensemble may otherwise be defined in terms of a different collection of rate functions $F'$, vector of parameters ${\vec{p}}\,'$, or environmental conditions restricting access to $\vec{x}^{0'}$ (\refsupp{}). We then ask what is the probability, given $A$ is a stable matrix, that $A'$ is also a stable matrix. This quantifies the intuitive statement that, over evolutionary timescales, it is not enough for a system to be stable, but rather that it have a reasonably high probability of remaining stable over continguous timeframes. In terms of a history $s^T$ over the states specifying the stability property, $\mathcal{S}$, this is given by
$$
r = \frac{\sum_{t=0}^{T-1} s_{t+1}^T s_{t}^T}{\sum_{t=0}^{T-1} s_{t}^T}.
$$
If the process $P(s^T)$ has limits $\mu(s) = \lim_{t,T \rightarrow \infty} P(s_t^T=s)$ and $\tau(s' \mid s) = \lim_{t,T \rightarrow \infty} P(s_{t+1}^T = s' \mid s_t^T = s)$ for $s,s' \in \{0,1\}$
, the expectation of $r$ as $T \rightarrow \infty$ is approximated by
\begin{equation}\label{eq:robustnessfromstoch}
R(S,\mu,\tau) = E[r] \approx \frac{\sum_{s,s'} \mu(s) \tau(s' \mid s)}{\sum_{s=1} \mu(s)}.
\end{equation}
This conditional probability corresponds to the parameter $R$ of the two-state process depicted in \ref{fig:robustnessprocess}A, and it is what we refer to as dynamical robustness.

In terms of Jacobian matrices $A$ and $A'$, if the underlying process $P([A]^T)$ has analogous limits
$$\mu(A) = \lim_{t,T \rightarrow \infty} P(A_t^T=A)$$
and
$$\tau(A' \mid A) = \lim_{t,T \rightarrow \infty} P(A_{t+1}^T = A' \mid A_t^T = A)$$
for $A,A' \in \mathbb{R}^{n \times n}$, \ref{eq:robustnessfromstoch} becomes
\begin{equation}
R(S,\mu,\tau) = \frac{\int_{A,A' \in \mathbb{R}^{n \times n}} d\mu(A) d\tau(A' \mid A) \mathcal{S}(A) \mathcal{S}(A')}{\int_{A \in \mathbb{R}^{n \times n}} d\mu(A) \mathcal{S}(A)}.
\end{equation}

In order to classify the properties of this process according to network architecture, we consider cases where the condition $\adj(G_A) = \adj(G_{A'})$ holds, which results in an analogous processes defined on each $G$-class $[G] \in \mathcal{F}/G$ given by
$$
P([A]^T \mid G_A {=} G) = \sum_{ \{ f^T \mid f_i^T \in [A]_i^T \} } P(f^T)
$$
for each history of length $T$, $[A]^T = ( [A_1], [A_2], \ldots, [A_T] ) \in [G]^T$
as visualized in \ref{fig:robustnessprocess}C. This now corresponds to a collection of ensembles of dynamical systems each with equivalent connectivities corresponding to a graph $G$. For each graph, there is a potentially different value of robustness given by
\begin{equation}\label{eq:robustnessbygraph}
\begin{aligned}
R(G,S,\mu,\tau)  =
 \frac{\int\limits_{A,A' \in \mathbb{R}^{n \times n} \atop G_A = G} d\mu(A) d\tau(A' \mid A) \mathcal{S}(A) \mathcal{S}(A')}{\int\limits_{A \in \mathbb{R}^{n \times n} \atop G_A = G} d\mu(A) \mathcal{S}(A)}.
\end{aligned}
\end{equation}
Comparing the values of $R$ for each $G$-class in $\mathcal{F}/G$ places a partial ordering on network architectures, $G$, which allows for the determination of which network architectures would be expected to be enriched relative to others in an evolutionary process where selection is imposed in a manner that results in a bias toward higher $R$ values.

\section{Abstracting network architectures to their strongly connected components and the symmetries of robustness}
The network architecture represented in terms of the adjacency matrix, \ref{eq:lininterdepadj}, can be abstracted into modules by mapping the interaction graph to the network of strongly connected components (SCCs). A SCC of a graph is a maximal subset of vertices where each vertex within the subset can be reached from any other \cite{Cormen2009}. The strongly connected components of some examples of three variable systems are outlined in \reffigscc{} along with their adjacency matrices.

\begin{figure*}[!ht]
\centering
\noindent\includegraphics[width=1.0\textwidth]{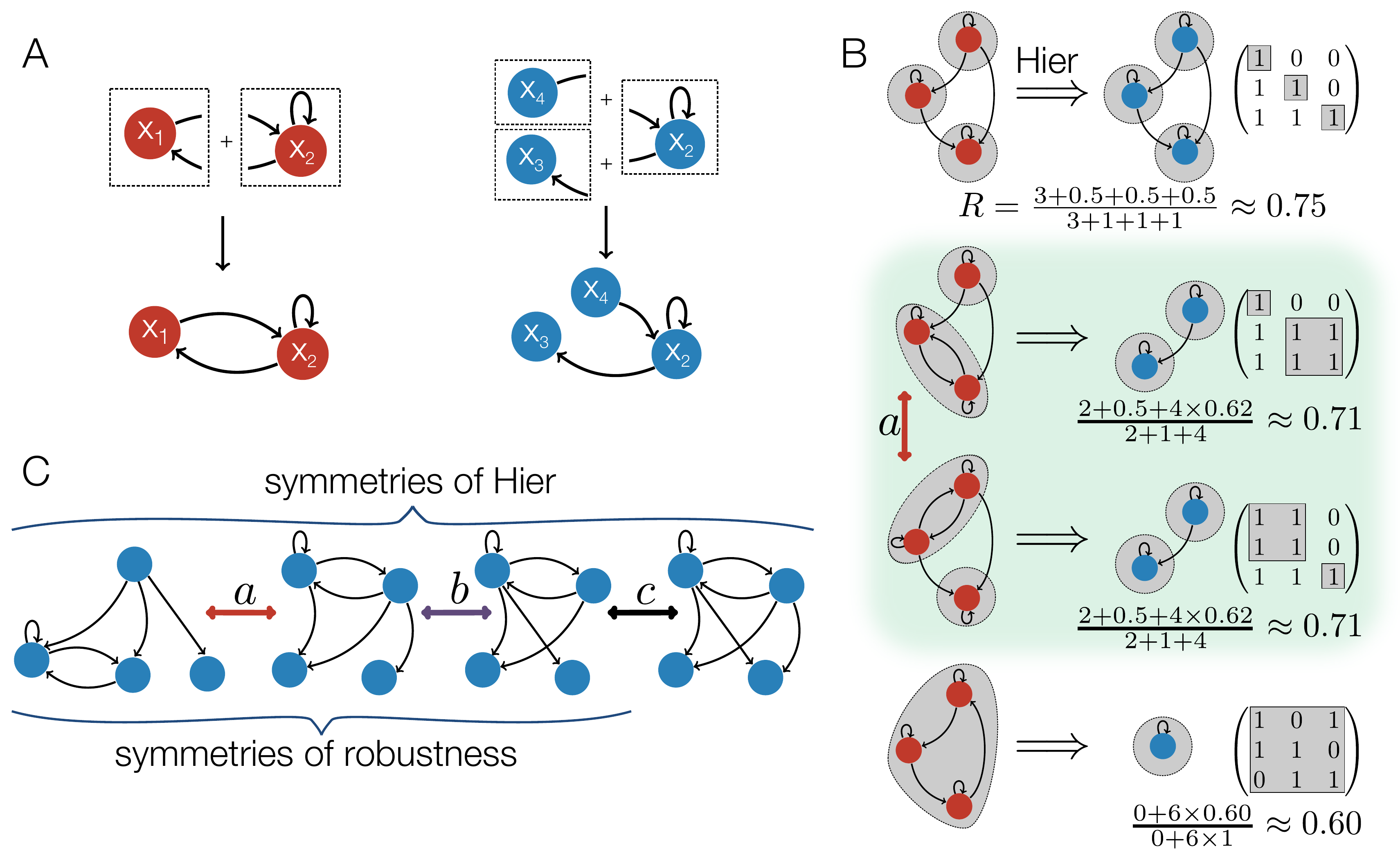}
\caption{{\bf Open systems, strongly connected components and symmetries of robustness.} (A) Example of the combination of open system modules to construct closed systems. (B) SCCs highlighted in gray for each of the four graphs representing the interdependencies relevant to four different three variable systems. The most hierarchical network, top panel, is the one that maximizes the number of SCCs and the number of links between them. We therefore define hierarchy as $max(\hbox{ED}) - \hbox{ED}$ where ED is the edit distance representing the number of link addition/deletion operations necessary to transform a given graph into the most hierarchical one. The two panels in the middle represent examples of hierarchical modular systems that posess both modularity (i.e. SCCs with more than one variable) and hierarchy. (C) Symmetries of the $\hier$ transformation between graphs and SCCs. The transformation $a$ represents an interchange of SCCs, $b$ moving a link between nodes in a component and $c$ adding a link. All three transformations represent symmetries of the $\hier$ transformation from graphs to SCCs while only $a$ and $b$ are symmetries of robustness.}
\label{fig:modsccsym}
\end{figure*}

The map from the interaction graph of a network to its SCCs, referred to as $\hier$ in \ref{fig:modsccsym}, results in a decomposition of $G$ into its SCCs. Each node of $\hier (G)$ corresponds to a strongly connected component of $G$. There is an edge from the node corresponding to component $C$ to the node corresponding to component $C'$ if and only if there exists a link from some vertex in $C$ to some vertex in $C'$ in $G$.  Because of the maximality of strongly connected components, $\hier (G)$ is acyclic.

One can also perform this construction in the opposite direction.  Start with a directed acyclic graph $H$.  To each node $n$ of $H$ associate
a strongly connected graph $C_n$.  To each link $(i,j)$ of $H$ associate a non-empty subset of $\Vertex(C_i) \times \Vertex(C_j)$.  The result will be a graph $G$ such that $\hier(G) = H$ and furthermore, every graph $G$ such that $\hier(G) = H$ can be obtained in this manner.

This map $\hier$ is many-to-one and so there is a large class of
operations which leaves $\hier(G)$ invariant for a given graph $G$ \reffighiertransformations{}. These three symmetries, \reffigscc{}, represent transformations that can be performed on the interaction graph that do not change the network of SCCs to which it is associated.
For instance, we may interchange the positions of the strongly
connected components relative to each other \reffighiertransformations$a$.  Leaving the components fixed, we may move links between nodes in a component \reffighiertransformations$b$ or between components, or even add or delete links \reffighiertransformations$c$.

\begin{figure}[!ht]
\centering
\noindent\includegraphics[width=0.9\columnwidth]{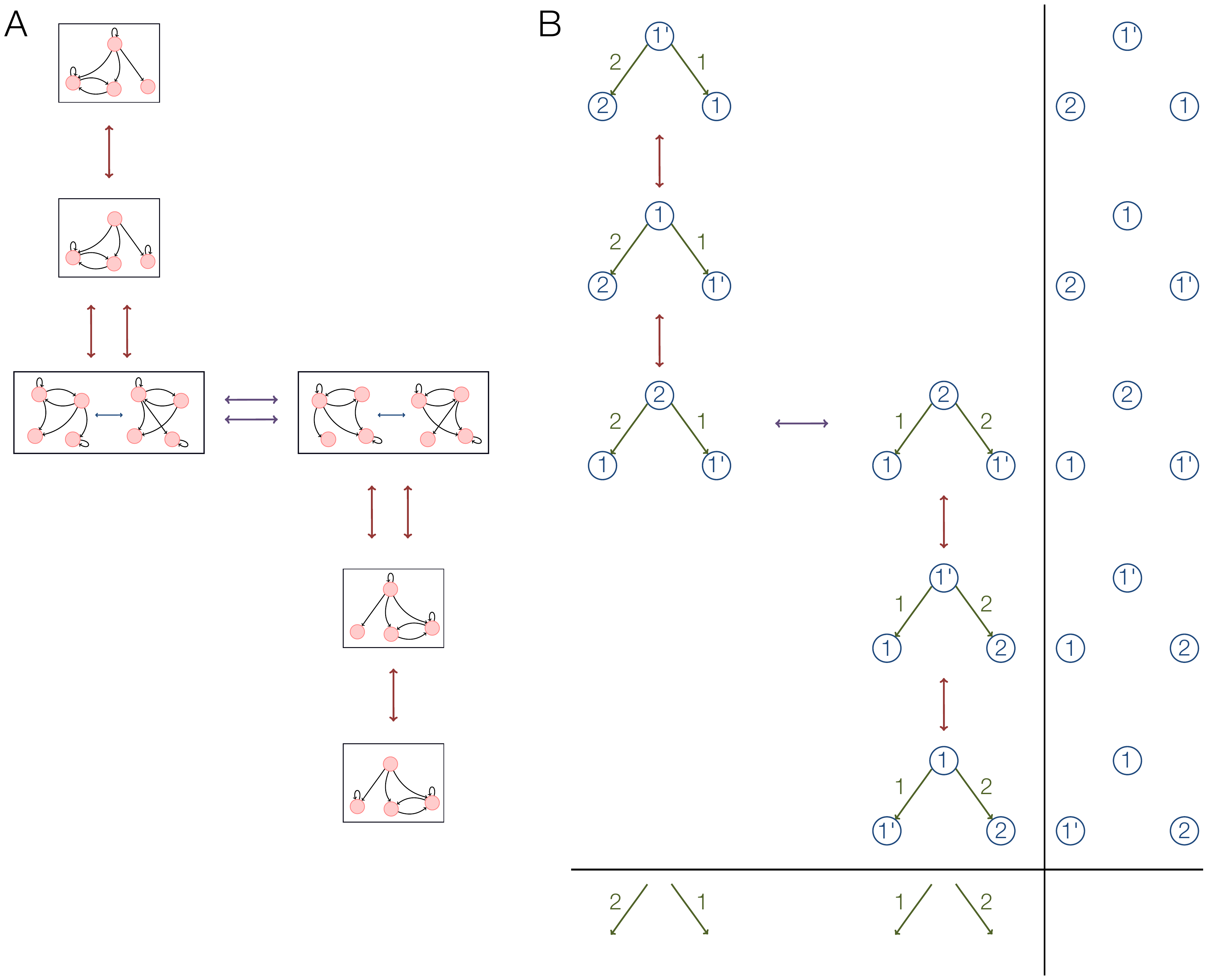}
\caption{{\bf Example symmetries of robustness.} In this example, the connected component sizes are fixed at $\{2,1,1\}$ with a total of $3$ links between them. Red arrows correspond to transformations like \reffighiertransformations a where SCCs are swapped whereas purple arrows correspond to transformations like \reffighiertransformations b where links are moved between nodes within a SCC. (A) shows all underlying graphs while (B) shows the number of nodes in each SCC and the number of links between the SCCs.}
\label{fig:robustnesssymmetries}
\end{figure}

Symmetries with respect to some property of the system are characterized by the ability to interchange these modules or their connectivity without changing that property. Two of these three intrinsic symmetries of $\hier$ are also symmetries with respect to dynamical robustness. \ref{fig:robustnesssymmetries} shows an example of these latter symmetries applied to a specific interaction graph.

\section{Derivation of the relationship between network architecture and robustness}
Now we derive an analytical expression for dynamical robustness, $R$, of a network in terms of its interaction graph, $G$, as a weighted average of the robustness, $R_{\alpha}$, of the SCCs, $C_{\alpha}$, the corresponding number of links within each SCC, $d_{\alpha}$, and the number of links between the SCCs, $l$ to give $d = l + \sum_{\alpha} d_{\alpha}$. The result ultimately holds for the case of simultaneously perturbing any number of elements of the Jacobian. To anchor the intuition before stating the more general result, we derive the expression for the case of perturbing a single element at each timestep of the process.
\subsection{Independent modification process to individual network interactions}
Let the index $i$ range over the non-zero entries $a_i$ of $A$. The entries of the Jacobian are sampled independently from a generic probability distribution $\rho_i$ for each entry. Under this assumption then
\begin{equation}\label{eq:singleresamplemutau}
\begin{aligned}
\mu(A) &= \prod_i \rho_i(a_i),\\
\tau^{(1)}(A' \mid A) &= \frac{1}{d} \sum_i \tau^{(1)}_i (A' \mid A),
\end{aligned}
\end{equation}
where
$$
\tau^{(1)}_i(A' \mid A) = \rho_i(a'_i) \prod_{j \neq i} \delta(a'_j - a_j).
$$

The decomposition of a digraph into SCCs corresponds to a block triangular decomposition of its adjacency matrix.  Say that the graph $G$ has SCCs $C_1, C_2, \ldots C_n$, which have been labelled in such a way that there are no links from vertices in component $C_i$ to component $C_j$ when $i < j$.  Label the vertices in such a way that $V_1, \ldots, V_{n_1}$ belong to $C_1$, $V_{n_1 + 1}, \ldots, V_{n_2}$ belong to $C_2$, etc.  Then, if we choose basis vectors corresponding to this labelling of the vertices, we will have $a_{ij} = 0$ whenever $i$ and $j$ correspond to different components and $i > j$.  This condition is equivalent to stating that the matrix is block triangular with blocks of size $n_1, n_2, \ldots$.

Since the determinant of a triangular matrix equals the product of the determinants of its diagonal blocks, it follows that the characteristic polynomial factors as the product of the charactericstic polynomials of its diagonal blocks.  Hence, a block triangular matrix is stable if and only if its diagonal blocks are stable.  Note that this condition does not depend upon the entries off the diagonal (which correspond to links between SCCs) and does not depend upon what order the components appear. This fact implies that the terms evaluating stability such as $\mathcal{S}(A)$ from \ref{eq:robustnessbygraph} decompose into products over the SCCs
\begin{equation}\label{eq:stabfactor}
\mathcal{S}(A) = \prod_{C_\alpha \in \hier(G)} \mathcal{S}(\pi_{C_{\alpha}}(A))
\end{equation}
where $\pi_{C_{\alpha}}(A)$ denotes the projection of the matrix $A$ onto the SCC $C_{\alpha}$.

\begin{figure}[!ht]
\centering
\noindent\includegraphics[width=0.9\columnwidth]{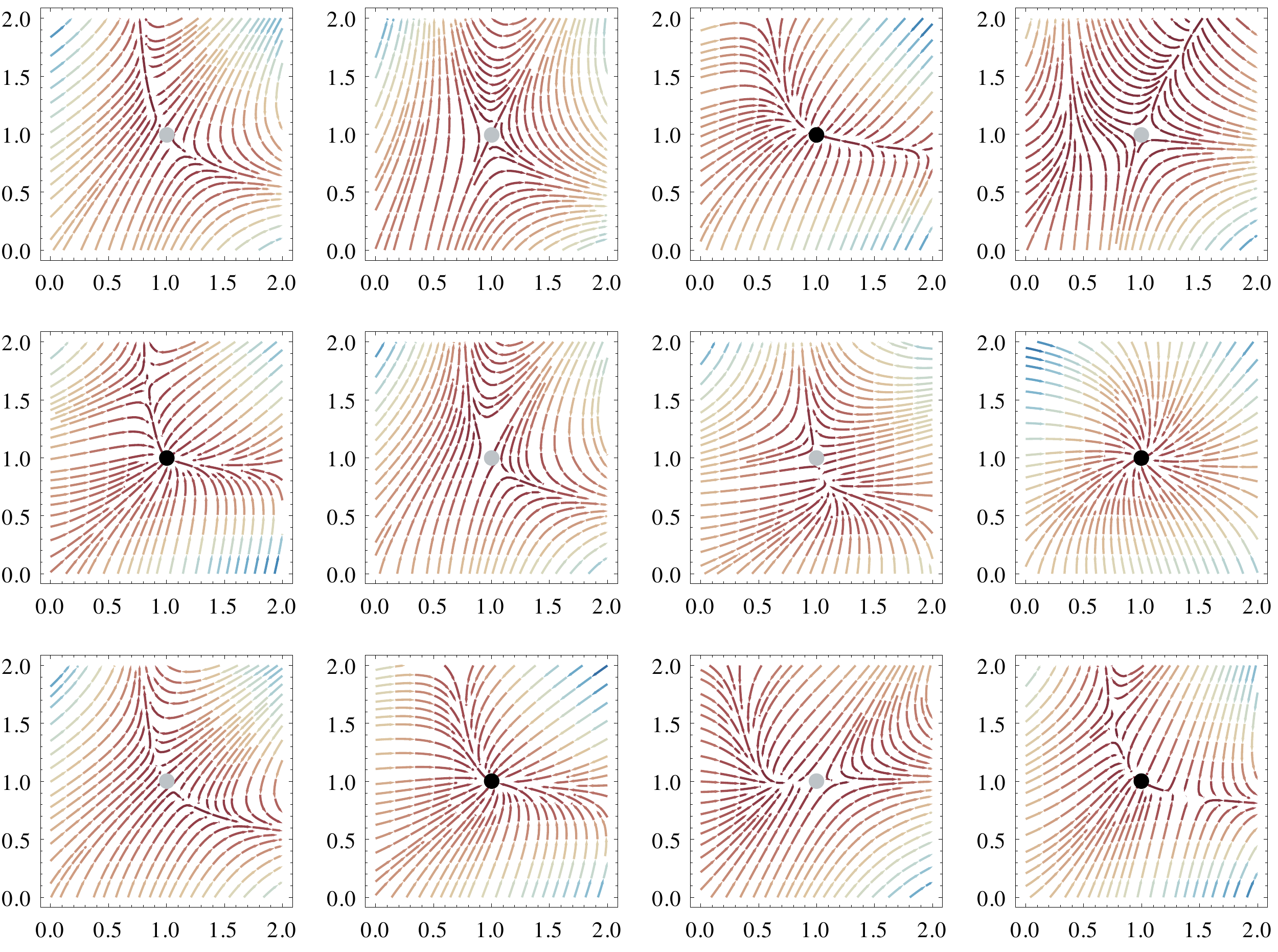}
\caption{{\bf Vector fields resulting from the random sampling of two component systems.} System parameters of \ref{eq:arbitraryjacobianreactionnetwork} are rescaled to ensure the fixed point is located at $(1,1)$. The color of the dot located at the fixed point indicates whether it is stable (black) or unstable (gray).}
\label{fig:jacobianvectorfields}
\end{figure}

To relate the robustness of a graph to the robustness of its SCCs, we substitute \ref{eq:singleresamplemutau} and \ref{eq:stabfactor} into \ref{eq:robustnessfromstoch}, collect factors corresponding to components, decompose integrals into their respective products, collapse integrals over delta distributions, and cancel common factors between numerator and denominator.  Let $L$ denote the set of edges of $G$ that connect distinct strongly connected components.  Then, if $i \in C_\alpha$ for some SCC $C_\alpha$, we have
\begin{widetext}
\begin{equation}\label{eq:robustncessforsccs}
\begin{aligned}
\frac{\int d\mu(A) d\tau^{(1)}_i (A' \mid A) \mathcal{S}(A) \mathcal{S}(A')}
       {\int d\mu(A) \mathcal{S}(A)}
&= \frac{\begin{matrix}
  \int \prod\limits_{k \in C_\alpha} da_k\,d{a'}_k\, \rho(a_k)
    \rho_i({a'}_i) \prod\limits_{j \in C_\alpha \setminus i} \delta ({a'}_j - a_j)
    \mathcal{S}(\pi_{C_{\alpha}}(A)) \mathcal{S}(\pi_{C_{\alpha}}(A')) \times \\
 \prod\limits_{C_\beta \in \hier(G) \setminus C_\alpha} \int
   \prod\limits_{j \in C_\beta} da_k\,d{a'}_k\, \rho({a'}_j) \delta ({a'}_j - a_j)
      \mathcal{S}(\pi_{C_{\beta}}(A)) \mathcal{S}(\pi_{C_{\beta}}(A')) \times \\
 \prod\limits_{l \in L} \int da_l\,d{a'}_l\,\rho({a'}_l) \delta({a'}_l - a_l) \end{matrix}}
{\begin{matrix}\int \prod\limits_{k \in C_\alpha} da_k\,\rho(a_k) \mathcal{S}(\pi_{C_{\alpha}}(A)) \times \\
 \prod_{C_\beta \in \hier(G) \setminus C_\alpha}
   \int \prod\limits_{h \in C_\alpha} da_h,\rho(a_h) \mathcal{S}(\pi_{C_{\beta}}(A)) \times \\
 \prod\limits_{l \in L} \int da_l\,\rho(a_l) \end{matrix}} \\
&= \frac{\int \prod\limits_{k \in C_\alpha} da_k\,d{a'}_k\,\rho(a_k)
    \rho_i({a'}_i) \prod\limits_{j \in C_\alpha \setminus i} \delta ({a'}_j - a_j)
    \mathcal{S}(\pi_{C_{\alpha}}(A)) \mathcal{S}(\pi_{C_{\alpha}}(A'))}
{\int \prod\limits_{k \in C_\alpha} dk\, \rho(a_k) \mathcal{S}(\pi_{C_{\alpha}}(A))} \\
&= R(C_{\alpha}, S, \mu_\alpha, \tau^{(1)}_\alpha)
\end{aligned}
\end{equation}\label{eq:robustncessforsccconnect}
\end{widetext}
where $\mu_\alpha$ and $\tau^{(1)}_\alpha$ refer to the analogues of \ref{eq:singleresamplemutau} for the subgraph $C_\alpha$ of $G$. \ref{eq:robustncessforsccs} shows that the terms in the sum over the elements of $A$, or equivalently the edges of $G$, that comprise a given SCC of $G$ reduce to the robustness of that SCC alone.  Likewise, when $i \in L$, we have
\begin{widetext}
\begin{equation}
\begin{aligned}
\frac{\int d\mu(A) d\tau^{(1)}_i (A' \mid A) \mathcal{S}(A) \mathcal{S}(A')}
       {\int d\mu(A) \mathcal{S}(A)}
&= \frac{\begin{matrix}
  \prod\limits_{C_\alpha \in \hier(G)} \int
   \prod\limits_{j \in C_\beta} da_k\,d{a'}_k\, \rho({a'}_j) \delta ({a'}_j - a_j)
      \mathcal{S}(\pi_{C_\alpha}(A)) \mathcal{S}(\pi_{C_\alpha}(A')) \times \\
 \int da_i d{a'}_i \rho_i(a) \rho_a({a'}_i) \times
 \prod\limits_{l \in L\setminus i} \int da_l\,d{a'}_l\,\rho({a'}_l) \delta({a'}_l - a_l) \end{matrix}}
{\prod_{C_\alpha \in \hier(G)}
   \int \prod\limits_{h \in C_\alpha} da_h,\rho(a_h) \mathcal{S}(\pi_{C_\alpha}(A)) \times
 \int da_i \rho(a_i) \times
 \prod\limits_{l \in L \setminus i} \int da_l\,\rho(a_l)} \\
&= 1
\end{aligned}
\end{equation}
\end{widetext}
For each $C_{\alpha} \in \hier(G)$, there will be $d_\alpha$ values of $i$ such that $i \in C_{\alpha}$ requiring instances of \ref{eq:robustncessforsccs}; likewise, there will be $l$ values of $i$ such that $i \in L$ requiring instances of \ref{eq:robustncessforsccconnect}.  Hence, when we perform the summation over $i$ to compute the robustness of $\mathcal{F}/G$ for each $G$, we will obtain a weighted average:
\begin{equation}\label{eq:robschematic}
R = \frac{l+d_1 R_1 + d_2 R_2 + \cdots}{l+d_1 + d_2 + \cdots}.
\end{equation}
Here $R_\alpha$ is shorthand for $R(C_\alpha,S,\mu_\alpha,\tau^{(1)}_\alpha)$ and $R$ is shorthand for $R(G,S,\mu,\tau^{(1)})$.
Examples of vector fields that correspond to the sampling of Jacobian matrices used in the computation of robustness for particular examples are shown in \ref{fig:robustnessconcept} and \ref{fig:jacobianvectorfields}. \ref{eq:robschematic} shows a schematized version of \ref{eq:robustnessbygraph} for the case of perturbing a single element at a time  (see \reffigscc$\,$ for examples demonstrating this expression). For instance, if our graph is the one in \reffigscc{} (middle panels), then we have two connected components, one with two nodes, and one with one node.  From \ref{tab:structstabmat}, we know that the graph with two nodes has probability $0.25$ of being stable and robustness $0.62$.  The graph with one node corresponds to a $1 \times 1$ matrix, so we have probability $0.5$ of stability and robustness $0.5$.  Thus, the probability of our 3-node graph being stable is $0.5 \times 0.25 = 0.125$ and its robustness is computed from \ref{eq:robschematic} in \reffigscc{}, which agrees with the value computed in \ref{tab:structstabmat} up to sampling error.

Examining this expression noting that $R_{\alpha}$ are all strictly less than one proves that networks maximizing $l$, will also maximize $R$.  Given two connected components $C_{\alpha}$ and $C_{\beta}$ with $v_{\alpha}$ and $v_{\beta}$ nodes respectively, we have a maximum of $v_{\alpha} v_{\beta}$ links going from $C_{\alpha}$ to $C_{\beta}$.  Hence, $l \le \sum_{(\alpha,\beta) \in \hier(G)} v_{\alpha} v_{\beta}$.  Since every acyclic digraph can be embedded into a totally ordered
set, we may assume without loss of generality that our components have
been ordered in a way such that, if $(\alpha,\beta) \in \hier(G)$, then $\alpha <
\beta$.  Hence, $l \le l_{max}$ where
$$l_{max} = \sum_{\alpha=1}^{n-1}\sum_{\beta=\alpha+1}^{n}v_{\alpha}
v_{\beta}~=~\frac{1}{2} \left( \sum_{\alpha=1}^{n} v_{\alpha} \right)^2-\frac{1}{2} \sum_{\alpha=1}^{n}
v_{\alpha}^2.$$
Suppose we have a graph $G$ with SCCs $C_1,\ldots,C_n$ and that $G_{\mathrm{tot}}$
is the graph on $n$ nodes with a link from node $i$ to node $j$ whenever $i < j$.  Then we have a graph $G_{\mathrm{max}}$ such that $\hier (G_{\mathrm{max}}) = G_{\mathrm{tot}}$, the components of $G_{\mathrm{max}}$ are also $C_1, \ldots, C_n$, but that includes all possible links between each pair of components. In this case $R(G)~\le~R(G_{\mathrm{max}})$ for all $G \ne G_{\mathrm{max}}$ where $R(G_{\mathrm{max}})$ is equivalent to \ref{eq:robschematic} with $l_{max}$ substituted for $l$
\begin{widetext}
\begin{align} \label{eq:sccmaxrobustness}
R(G_{\mathrm{max}}) =
\frac{\frac{1}{2} ((\sum_{\alpha=1}^n v_{\alpha})^2 - \sum_{\alpha=1}^n v_{\alpha}^2) +
                    \sum_{\alpha=1}^n \connectivity_{\alpha}
                                    R(C_{\alpha},S,\mu_{\alpha},\tau_{\alpha})}
     {\frac{1}{2} ((\sum_{\alpha=1}^n v_{\alpha})^2 - \sum_{\alpha=1}^n v_{\alpha}^2) +
                    \sum_{\alpha=1}^n \connectivity_{\alpha}}.
\end{align}
\end{widetext}
\subsection{Modifying multiple interactions simultaneously}
This argument also works when relationships between multiple system components are perturbed simultaneously,
although the notation becomes more complicated.  Suppose that we
resample $m$ interactions.  Then the analogue of \ref{eq:singleresamplemutau} is
\begin{equation}\label{eq:multipleresamplemutau}
\begin{aligned}
\tau^{(m)}(A' \mid A) &= \frac{1}{\binom{d}{m}} \sum_{i_1, \ldots, i_m}
  \tau^{(m)}_{i_1, \ldots, i_m} (A' \mid A), \\
\tau^{(m)}_{i_1, \ldots, i_m} (A' \mid A) &=
 \prod_{k=1}^m \rho_{i_k} (a'_{i_k})
 \prod_{j \notin \{i_1, \ldots i_m\}} \delta(a'_j - a_j).
\end{aligned}
\end{equation}
Now define
\begin{widetext}
\begin{equation*}
M = \left\{(m_0, m_1, \ldots, m_n) \,\bigg|\,
m = \sum_{i=0}^n m_i \quad\&\quad
m_0 \le \sum_{i=1}^{n-1} \sum_{j=i+1}^n \ell_{ij} \quad\&\quad
(\forall i \in \{1, \ldots, n\}) \; m_i < \connectivity_i \right\}.
\end{equation*}
\end{widetext}
Then, given $(m_0, m_1, \ldots, m_n) \in M$, there are ${m \choose
m_0, m_1, \ldots. m_n}$ ways of choosing $m_i$ links from $C_i$ and
$m_0$ links between strongly connected components.  Hence, our
weighted average becomes
\begin{widetext}
\begin{equation}\label{eq:robustnessmultiple}
R(G, S, \mu, \tau^{(m)}) =
\frac{\sum\limits_{(m_0, m_1, \ldots m_n) \in M}
      {m \choose m_0, m_1, \ldots. m_n}
      \left(m_0 + \sum\limits_{\alpha=1}^n m_\alpha R(C_\alpha, S, \mu, \tau^{(m_\alpha)}) \right)}
     {\sum\limits_{(m_0, m_1, \ldots m_n) \in M}
      {m \choose m_0, m_1, \ldots. m_n} m}
\end{equation}
\end{widetext}

As before, since $R(C_\alpha, S, \mu, \tau^{(m_i)}) \le 1$, we may
increase $R(G, S, \mu, \tau^{(m)})$ by increasing the maximum
possible value of $m_0$ while keeping the strongly connected
components the same.  Again, if we fix $\hier(G)$, the maximum
possible value of $m_0$ is $\sum_{(\alpha,\beta) \in \hier(G)} v_\alpha v_\beta$
whereas, if we allow it to vary, the maximum is $\frac{1}{2} ((\sum_{\alpha=1}^n
v_\alpha)^2 - \sum_{\alpha=1}^n v_\alpha^2)$, which is attained when $\hier(G_{max}) =
G_{tot}$.  Hence, we conclude that $R(G, S, \mu, \tau^{(m)}) \le
R(G_{\mathrm{max}}, S, \mu, \tau^{(m)})$.

This implies that the interaction graphs for systems that are the most robust will maximize the number of links between SCCs as well as the overall number of SCCs with respect to a particular system size. This analytical result predicts that any network whose associated dynamical system has the interaction graph equivalent to the total ordering will be more robust than those associated to any of the other interaction graphs in \reffigscc{}. The graph associated to the total ordering is the most hierarchical network architecture for any given number of system components like that of \reffigscc{} top for three component systems where the \emph{highest} component in the hierarchy has directed links to all other nodes in the network, the second highest component has directed links to all other nodes in the network except the highest one, \emph{et cetera} (\refsupp{}). Because this result is purely topological in nature, it does not depend at all upon any particular details such as the probability distribution from which the component interaction strengths are sampled or the size of the system. The result that dynamical robustness is correlated with network hierarchy therefore applies to an even broader class of dynamical systems than the particular random ensembles we have studied directly.

To test the prediction of the analytical results in \ref{eq:robschematic} and \ref{eq:robustnessmultiple}, we computed approximations to the probability distribution of stability and dynamical robustness relative to network architecture for ensembles of systems having two or three interacting components (see \refsupp{} \ref{tab:structstabmat} and \ref{tab:structstabmat3}). For all of these, we found that robustness is correlated with connectivity, but that the most robust systems have intermediate connectivity for a given network size (\reffigrobustconnect). Accounting for the number of cycles in a network architecture reveals a strong correlation between robustness and connectivity that was hidden when networks with any number of cycles were considered together (\reffigconnectcycle3D3x3). While the most hierarchical network architecture will always lack cycles altogether, cycle number alone is clearly insufficient to account for robustness as the members of each class span nearly the entire range of possible robustness values. Consistent with our analysis of the symmetries of robustness, we found that the most hierarchical network architecture is the most robust (\reffigrobusthierarchy). Moreover, if we consider hierarchy partitioned by connectivity, we find that there is a monotonic increase in robustness following any line of increasing hierarchy in \reffigconnectdist3D3x3.

\begin{figure*}[!ht]
\centering
\noindent\includegraphics[width=1.0\textwidth]{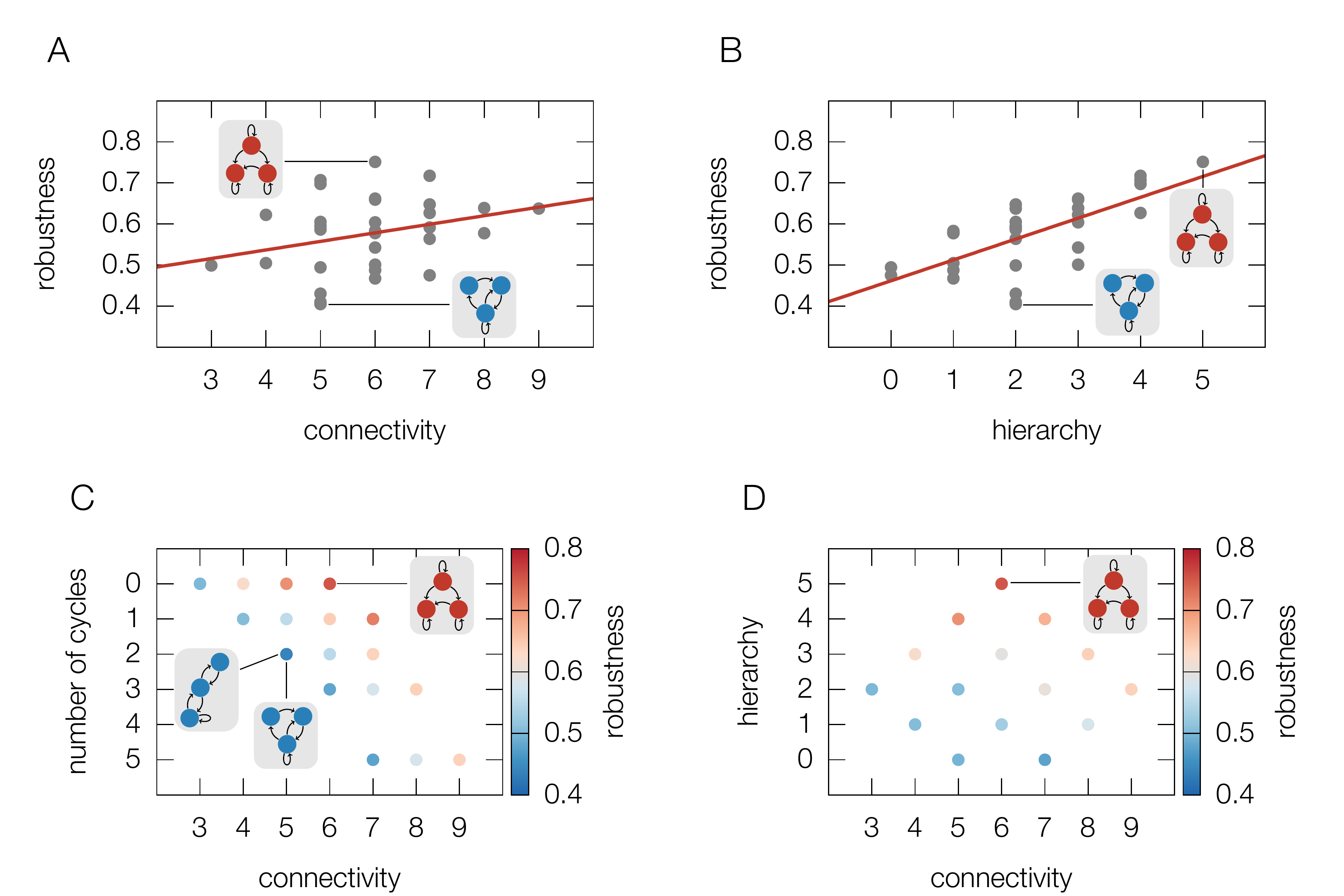}
\caption{{\bf Characterization of stability and robustness according to properties of system structure for three variable systems} (A) Robustness versus connectivity. The red line represents a best fit in the least-squares sense with Pearson product-moment correlation coefficient $r=0.29$. The lowest and highest robustness network architectures are labelled. Other network architectures are shown in \ref{tab:structstabmat3}. (B) Robustness versus hierarchy. Correlation coefficient $r=0.67$. (C) Number of cycles and (D) hierarchy vs connectivity and robustness. The color of each point represents the average robustness of all graphs having the parameters specified on the $x$ and $y$ axes.
}
\label{fig:combined}
\end{figure*}

\section{Conclusion}
Our analysis predicts that, in general, an ensemble of systems where robustness has been the predominant object of selection and has been positively selected over a sufficiently long period of time should exhibit a bias toward more hierarchical network topologies. Given the manner in which we define robustness in \ref{eq:robustnessfromstoch}, this is a very general constraint. In the short term, this prediction may be further evaluated at the levels of both metabolic and transcription factor networks, which have already been shown to display hierarchical structure, but whose dynamics have not been sufficiently well characterized to ascertain their dynamical robustness as we have defined it here \cite{Zhao2006,Bhardwaj2010,Colm}. At the ecological level, a system subjected to the environmental stress of overfishing, which may imply selection for robustness, has been observed to exhibit such a bias toward more hierarchical network architectures \cite{Bascompte2005}.
In the long term, this prediction may be evaluated using experimental evolution by comparing the degree of hierarchy that emerges in the evolution of gene regulatory network topology in the context of both static and fluctuating environments that impose differential selection strengths for dynamical robustness \cite{Leroi1994}.

In order to further this work from a theoretical perspective, it will be necessary to deepen our understanding of the relationship between dynamical robustness and the underlying network topology. Following May \cite{Cohen1984,May1972a,Radius2014,Majumdar2014}, this will involve improving the general understanding of the relationship between perturbations to a system's structure and the qualitative changes in the dynamical phenomena it can produce. The conservation of robustness with respect to nontrivial symmetries including the interchange of SCCs and permutation within SCCs suggests the existence of an evolutionary neutral space. A deeper mathematical characterization of the full symmetry groupoid of dynamical robustness may thus help to characterize this potential evolutionary constraint \cite{Golubitsky2006}.
For some classes of systems, it may be possible to go beyond the linear approximation and corresponding local summary statistics of the phase space, such as dynamical robustness, to provide a more complete characterization of the relationship between network architecture and the global structure of the phase space of the corresponding ensemble of biological networks.

The relationship between structure and function is fundamental to networks at every level of the biological hierarchy. Equally fundamental is the ability of systems to persist over long periods of time, which is dependent upon their dynamical robustness. Here we have demonstrated a structure-function relationship wherein biological networks that are more hierarchical are more robust and thus more likely to persist when this feature is the dominant object of selection in the evolutionary process.

\section*{Acknowlegements}
Support was provided by NIH MSTP training grant T32-GM007288 to CS and NIH R01-CA164468-01 and R01-DA033788 to AB. The authors would like to thank Ximo Pechuan, Daniel Biro, and Jay Sulzberger for helpful suggestions and discussion.

\bibliographystyle{pnas}
\bibliography{bib/books,bib/papers}

\pagebreak
\FloatBarrier

\beginsupplement
\setcounter{secnumdepth}{4}

\begin{widetext}
\begin{center}
\text{\large \refsupp{} for }\\
\textbf{\large Hierarchical Network Structure Promotes Dynamical Robustness}\\
\text{Cameron Smith, Raymond S. Puzio, Aviv Bergman$^*$}\\
\text{$^*$Corresponding author. E-mail: aviv@einstein.yu.edu}
\end{center}
\end{widetext}

\section{Stability and robustness analysis of particular system ensembles}

Here we compute robustness values for particular examples, where we choose the distributions $\rho_i$ to all be the uniform distribution $\mathcal{U}(-1,1)$ on the $\connectivity$-dimensional hypercube, $H^\connectivity$, of edge length $r=2$, centered about the origin.  For this choice, we will have $\rho_i = \mathbf{1}_{[-1,1]}$.

For systems having two variables, we can analytically compute the probability of stability and robustness from \ref{eq:robustnessbygraph}. For those having three variables, we can estimate these same quantities using Monte Carlo simulations. Systems of larger size can be analyzed using the symmetry properties of robustness extracted from this analysis. We note again that while we use the uniform distribution for the purposes of illustration, the analysis could be performed for other distributions and our result relating network hierarchy to robustness in \ref{eq:robschematic} and \ref{eq:robustnessmultiple} is independent of the form of this distribution. For two-variable systems having $2 \times 2$ Jacobian matrices, the aforementioned stability criteria result in the conditions $T < 0$ and $D >
0$ where $T$ and $D$ denote the trace and the determinant. Suppose we have a stable matrix
$$
\begin{bmatrix}
a & b \\
d & c
\end{bmatrix}
$$
where $a + c < 0$ and $ac > bd$.  For the case in which $x_1=a,\,x_2=b,\,x_3=c,\,x_4=d$ we need to compute what corresponds to $R(G,S,\mu,\tau^{(1)}_k)$ where $k=1 \ldots 4$. By symmetry, there are two cases to consider; resampling $a$ is equivalent to resampling $c$ and resampling $b$ is equivalent to resampling $d$ so we only need to explicitly compute $R(G,S,\mu,\tau^{(1)}_1)$ and $R(G,S,\mu,\tau^{(1)}_2)$. Suppose that we resample $b$ to compute $R(G,S,\mu,\tau^{(1)}_2)$.
The denominator of \ref{eq:robustnessbygraph} in this case is given by
\begin{align*}
P\left(\mathcal{S} \left( \begin{bmatrix}
a & b \\
d & c
\end{bmatrix} \right) = 1 \right) = \frac{\int_{\genfrac{}{}{0pt}{}{\genfrac{}{}{0pt}{}{ac>bd}{a+c<0}}{H^4}} da\,db\,dc\,dd\,1}{\int_{H^4} da\,db\,dc\,dd\,1}.
\end{align*}
Since the trace does not involve $b$, the $T<0$ condition will be satisfied automatically and we only need to examine the determinant. Thus, we have the inequalities $ac > b'd$ and $-1 < b' < 1$ in addition to the previous constraints leading to an expression for the numerator of \ref{eq:robustnessbygraph}
\begin{align*}
& P\left(\mathcal{S} \left( \begin{bmatrix}
a & b \\
d & c
\end{bmatrix} \right)=1 \textrm{ and } \mathcal{S} \left( \begin{bmatrix}
a & b' \\
d & c
\end{bmatrix} \right)=1 \right) = \\
& \frac{\int_{{{ac>b'd \atop ac>bd} \atop a+c<0} \atop H^5} da\,db\,dc\,dd\,db'\,1}{\int_{H^5} da\,db\,dc\,dd\,db'\,1}.
\end{align*}
The analogous equation for resampling $a$ is
\begin{align*}
& P\left(\mathcal{S} \left( \begin{bmatrix}
a & b \\
d & c
\end{bmatrix} \right)=1 \textrm{ and } \mathcal{S}
\left( \begin{bmatrix}
a' & b \\
d & c
\end{bmatrix} \right)=1 \right) = \\
& \frac{\int_{{{{a'c>bd \atop a' + c < 0} \atop ac>bd} \atop a+c<0} \atop H^5} da\,db\,dc\,dd\,da'\,1}{\int_{H^5} da\,db\,dc\,dd\,da'\,1}.
\end{align*}
Using this approach the probability of stability and of robustness for all two variable systems is given in \ref{tab:structstabmat}.

The analogous results for all three variable systems are computed using Monte Carlo integration and shown in \ref{tab:structstabmat3} and \reffigrobustconnect. This process is associated with some error relative to the exact integration described above. In all simulations we use $N~=~10000$ so that the maximum error is $0.005$ (see \ref{suppsec:montecarlo}).

\begin{figure}[!ht]
\centering
\noindent\includegraphics[width=0.6\columnwidth]{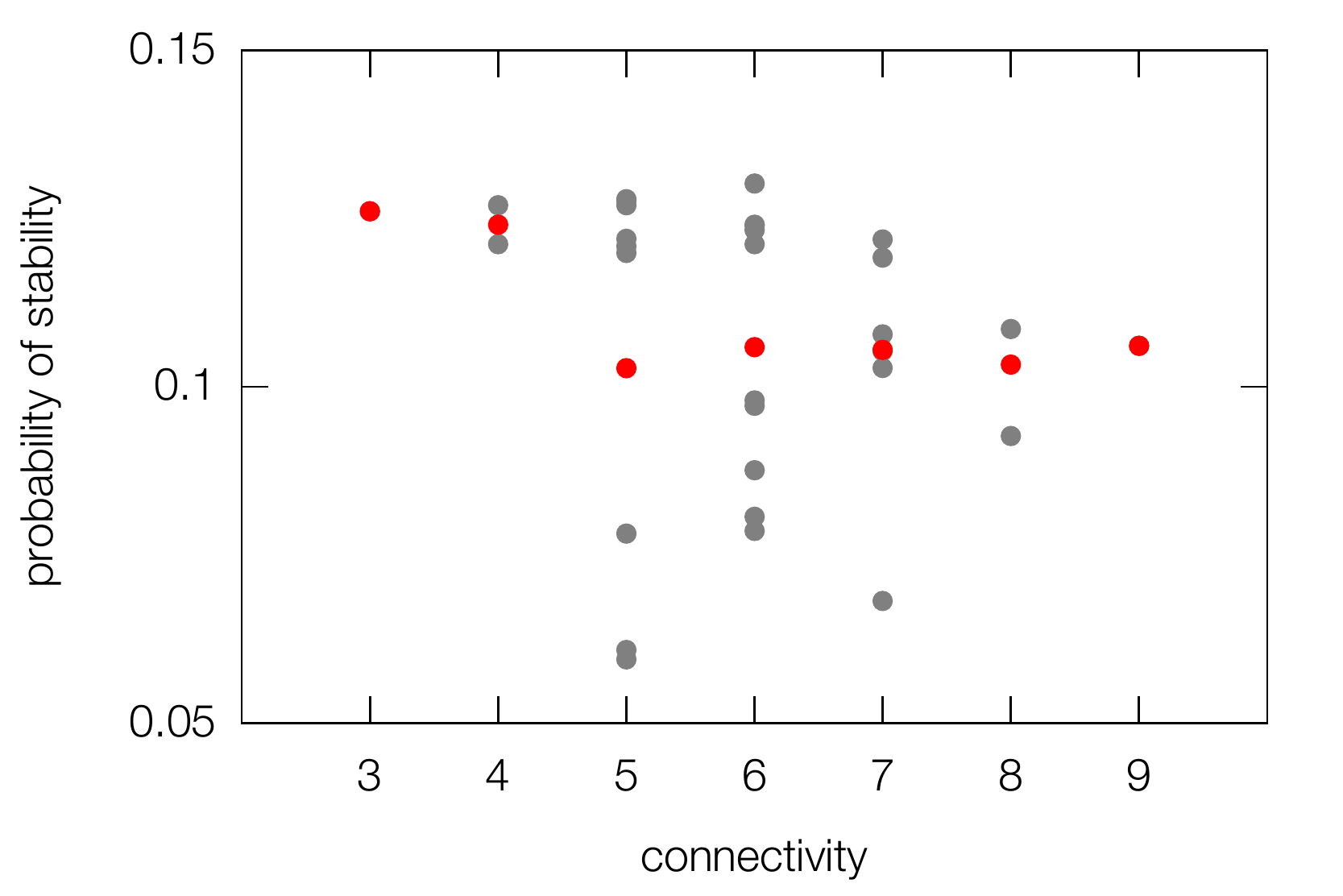}
\caption{{\bf System stability as a function of connectivity.} The red points represent the average of system stability at each connectivity.}
\label{fig:apstab3x3}
\end{figure}

It has been stated previously on the basis of simulation that system stability decreases with connectivity as the system size goes to infinity \cite{May1972}. For small system sizes such as the two and three variable systems, the situation is not so clear cut. For two variable systems, system stability is constant across the entire range of connectivities. For three variable systems, the trend shows a minor decrease from connectivity $4$ to $5$ followed by small fluctuations as shown in \ref{fig:apstab3x3}.

\begin{figure}[!ht]
\centering
\noindent\includegraphics[width=0.6\columnwidth]{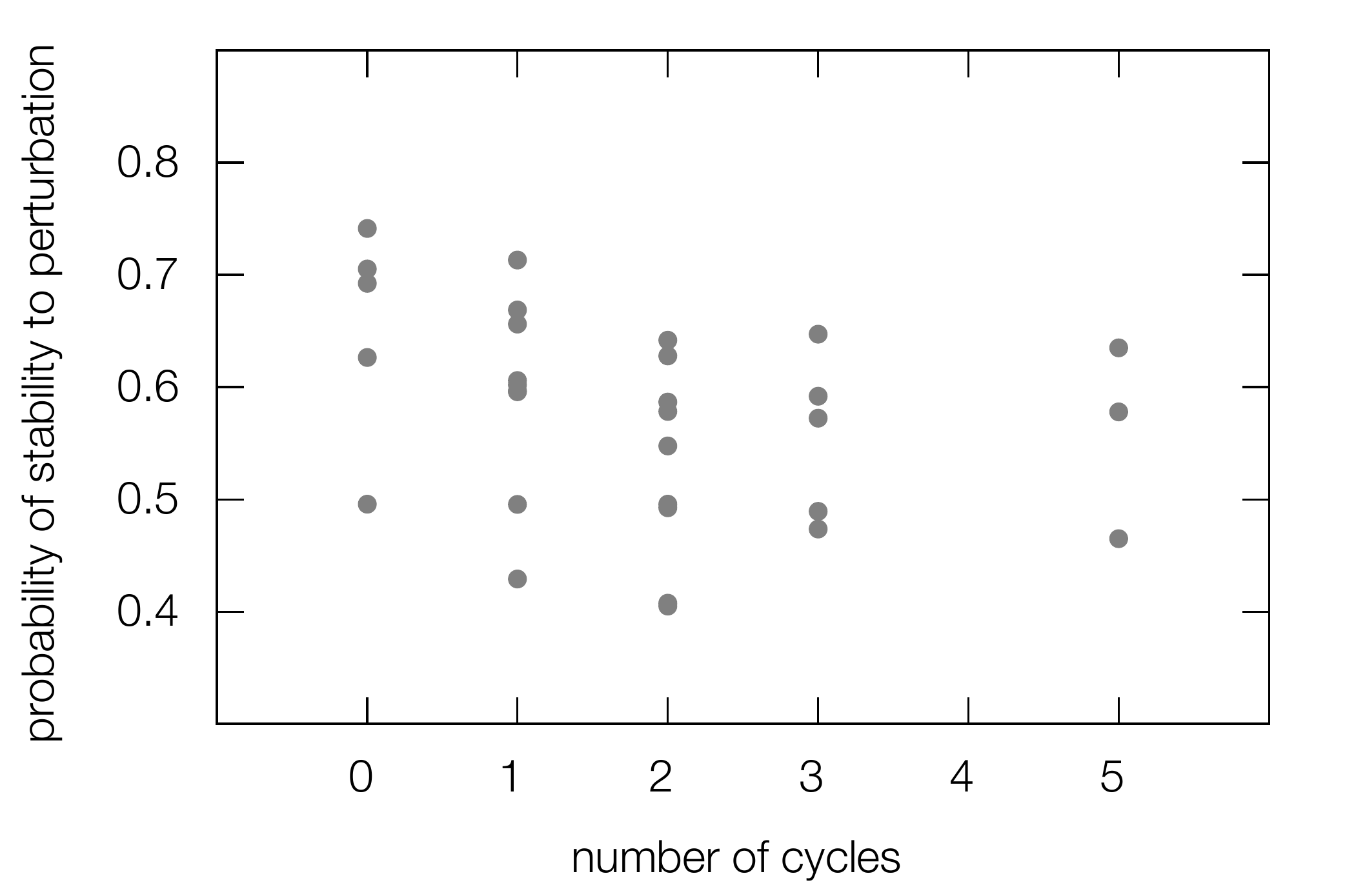}
\caption{{\bf dynamical robustness as a function of number of cycles.} }
\label{fig:cycle3x3}
\end{figure}

The relationship between connectivity and robustness for two variable systems is shown in \ref{tab:structstabmat} and likewise for three variable systems in \ref{tab:structstabmat3} and \reffigrobustconnect. If we average over the different classes of matrices for a given connectivity we see there is a correlation between connectivity and robustness demonstrated by the red lines in \reffigrobustconnect.
\ref{fig:cycle3x3} shows the robustness for all three variable systems as a function of the number of simple cycles (elementary circuits) of length greater than one in the corresponding directed graph \cite{Johnson1975}. There appears to be a weak negative correlation between robustness and the number of simple cycles.

The combination of connectivity and cycle number as shown in \reffigconnectcycle3D3x3 provides a better classification of the dependence of robustness upon network topology. Here the robustness of three variable systems with a given number of cycles, increases monotonically with connectivity. The network with the highest robustness for three variable systems is that of \reffigscc{} (top panel). This network is the most hierarchical of all three variable systems in the sense that it represents a total ordering of the components of the network and its adjacency matrix also shown in \reffigscc{} (top panel) has a block triangular structure.

This observation suggested that graph edit distance from \reffigscc{} (top panel), hierarchy, might provide a better characterization of dynamical robustness. \reffigrobusthierarchy$\,$ shows dynamical robustness as a function of hierarchy. There is a monotonic correlation between the upper bound of robustness and hierarchy. \reffigconnectdist3D3x3$\,$ shows dynamical robustness as a function of both hierarchy and connectivity. The monotonic correlation between hierarchy and robustness is refined by an underlying correlation between robustness and connectivity analogous to that of \reffigconnectcycle3D3x3.

\section{Monte carlo integration}\label{suppsec:montecarlo}
If we sample $N = N_{\mathrm{stab}} + N_{\mathrm{unstab}}$ matrices where each has some probability $\theta$ of being stable then $N_{\mathrm{stab}}$ has a binomial distribution. We can compute a sample estimate for $\theta$, $\hat{\theta} = \frac{N_{\mathrm{stab}}}{N}$ \cite{Murphy2012}. The posterior distribution in this case is known to be a Beta distribution as a result of Beta-Binomial conjugacy
$$
\mathrm{Beta}(\theta | \mathcal{D}) = \mathrm{Beta}(\theta | N_{\mathrm{stab}} + a, N_{\mathrm{unstab}} + b)
$$
where $a$ and $b$ are the hyperparameters of the Beta prior and we consider the uninformative uniform prior corresponding to $a=b=1$. We consider the maximum a posteriori estimate
$$\hat{\theta}_{MAP} = \frac{a + N_{\mathrm{stab}} - 1}{a + b + N - 2}$$
which corresponds in this case to the maximum likelihood estimate
$$
\hat{\theta}_{MLE} = \frac{N_{\mathrm{stab}}}{N}.
$$
This estimate is characterized by the variance of the posterior Beta distribution
\begin{align*}
&\mathrm{var}(\theta | \mathcal{D}) =\\
&\frac{(a+N_{\mathrm{stab}})(b+N_{\mathrm{unstab}})}{(a + N_{\mathrm{stab}} + b + N_{\mathrm{unstab}})^2(a + N_{\mathrm{stab}} + b + N_{\mathrm{unstab}}+1)}
\end{align*}
Since for the chosen prior $a=b=1 \ll N$ this simplifies to
$$
\mathrm{var}(\theta | \mathcal{D}) = \frac{\hat{\theta}(1-\hat{\theta})}{N}
$$
yielding the error estimate given by the associated standard deviation. In all simulations we use $N~=~10000$ so that the maximum error for $\hat{\theta}~=~0.5$ is $\sigma~=~\sqrt{\mathrm{var}(\theta | \mathcal{D})} \approx 0.005$.

\section{Reaction Networks, Gene Regulatory Networks, and Ecological Networks with Prescribed Connectivity and Jacobians}\label{sec:reactionnetjacobian}

The quality of interest in this paper is robustness, which is related to the concept of structural stability \cite{Smale1967}, whose evaluation requires the determination of whether or not a given dynamical system that is determined to be stable remains stable under a perturbation to one or more of its defining parameters, its rate functions, or environmental constraints that restrict it to a subset of its basins of attraction. We mean to refer to perturbations to the structure of the system itself as determined by the strengths of the couplings between the components and not only to perturbations of the state vector at a given point in time. It is justified to consider resampling elements of $A$ to generate $A'$ as a proxy for resampling elements of $\vec{p}$ to produce $\vec{p}\,'$ if any matrix $A$ can be obtained for some $F_i$, $\vec{p}$ and $\vec{x}^0$. This holds for the $F_i$ defining the Lotka-Volterra model. This is due to the fact that for a specification of non-zero real numbers for the components of $\vec{n}^0$ and any real numbers for the components of $a_{ij}$, there is a choice of parameters $\vec{p}$ given by $b_{ij} = \frac{a_{ij}}{n_i^0}$ and $r_i = - \sum_{j=1}^N \frac{n_j^0}{n_i^0} a_{ij}$ that generates those particular $a_{ij}$ as the Jacobian matrix of the dynamical system. Checking this property of the domain of realizability of the Jacobian can be done for ensembles of systems other than the Lotka-Volterra ensemble. For arbitrary biochemical reaction and gene regulatory networks, this property is likely to hold so long as not too many types of transformations are constrained from possibility. For example, a simplified version of the general form of the gene regulatory network model presented in \ref{fig:biomodelexamples} center panel is given by the system
\begin{equation}
\frac{dg_i}{dt} = \sum_{j=1}^N k_{ij} g_j,
\end{equation}
with one parameter $k_{ij} \in \mathbb{R}$ for every pair $(g_i,g_j)$ of genes. The Jacobian of this system is $A_{ij} = k_{ij}$, and, therefore, sampling parameters of the model is precisely equivalent to sampling elements of the Jacobian.

To justify our consideration of arbitrary Jacobian matrices in the case of reaction networks, we determine a simple ensemble for which arbitrary Jacobian matrices are realizable. This condition holds if one can solve for the parameter values of the system of equations corresponding to that ensemble in terms of the elements of an arbitrary Jacobian matrix. More precisely, we will show that, given an arbitrary directed graph $G$ where $G_{ii} = 1$ for all $i$, there exists a system of reactions having $G$ as its interaction graph and satisfying the following property: For any point $\vec{x}^0$ in the positive orthant and an arbitrary matrix $M$ whose interaction graph is $G$, there exists a choice of non-negative rates such that $\vec{x}^0$ is a fixed point of the network and the Jacobian equals $M$ at $\vec{x}^0$.

We begin by noting that, since the form of the rate equations for reaction networks are invariant under rescaling the concentrations and rate constants, we can make the coordinates of the point $\vec{x}^0$ be $(1,1,\ldots,1)$.  This will simplify the computation.

Let $N$ be the number of nodes of $G$.  Our reaction net will consist of $N$ species of reactants, $A_1, \ldots, A_N$, whose concentrations are $c_1, \ldots, c_N$.  The reactions are defined as follows:
\begin{equation}\label{eq:arbitraryjacobianreactionnetwork}
\begin{aligned}
\emptyset &\to A_i, & 1 \le i \le N,\\
A_i &\to \emptyset, & 1 \le i \le N,\\
2A_i &\to 3A_i, & 1 \le i \le N,\\
A_i + A_j &\leftrightarrow A_j, & i \neq j,\, 1 \le i,j \le N,\, G_{ij} = 1.
\end{aligned}
\end{equation}

The rate equations for such a system are:
\begin{equation}\label{eq:crnarbitraryjacobian}
\begin{aligned}
\frac{dc_i}{dt} = &F_i = k_{\emptyset \to A_i} - k_{A_i \to \emptyset} c_i + k_{2A_i \to 3A_i} c_i^2 \\
&+ \sum_{\substack{1 \le j \le N \\ j \neq i \\ G_{ij} = 1}} k_{A_j \to A_i + A_j} c_j - k_{A_i + A_j \to A_j} c_i c_j
\end{aligned}
\end{equation}
The Jacobian at $\vec{x}^0$ is given as
\begin{align*}
\left. \frac{\partial F_i}{\partial c_i}\right|_{\vec{x}^0} &= - k_{A_i \to \emptyset} + 2 k_{2A_i \to 3A_i} - \sum_{\substack{1 \le j \le N \\ j \neq i \\ G_{ij} = 1}} k_{A_i + A_j \rightarrow A_j}, & \\
\left. \frac{\partial F_i}{\partial c_j}\right|_{\vec{x}^0} &= k_{A_j \to A_i + A_j} - k_{A_i + A_j \to A_j}, &
\end{align*}
where $i \neq j$.
By combining the equations $F_i(\vec{x}^0) = 0$ from \ref{eq:crnarbitraryjacobian} and $\frac{\partial F_i}{\partial c_j}|_{\vec{x}^0} = M_{ij}$ we obtain the equivalent system of equations
\begin{align}
& k_{2A_i \to 3A_i} - k_{\emptyset \to A_i} = M_{ii} + \sum_{\substack{1 \le j \le N \\ j \neq i \\ G_{ij} = 1}} k_{A_j \to A_i + A_j} \label{eq:jacobianconstraint1}  \\
& 2k_{2A_i \to 3A_i} - k_{A_i \to \emptyset} = M_{ii} + \sum_{\substack{1 \le j \le N \\ j \neq i \\ G_{ij} = 1}} k_{A_i + A_j \to A_j} \label{eq:jacobianconstraint2}\\
& k_{A_j \to A_i + A_j} - k_{A_i + A_j \to A_j} = M_{ij} \label{eq:jacobianconstraint3}
\end{align}
We may solve these equations for the rate constants as follows.  We begin by solving \ref{eq:jacobianconstraint3} by either choosing $k_{A_i + A_j \to A_j} \ge 0$ and setting $k_{A_j \to A_i + A_j} = M_{ij} + k_{A_i + A_j \to A_j}$ when $M_{ij} \ge 0$ or choosing $k_{A_j \to A_i + A_j} \ge 0$ and setting $k_{A_i + A_j \to A_j} = k_{A_j \to A_i + A_j} - M_{ij}$ when $M_{ij} < 0$.  Pick
\begin{equation}
\begin{aligned}
k_{2A_i \to 3A_i} \ge \max \bigg(&0, M_{ii} + \sum_{\substack{1 \le j \le N \\ j \neq i \\ G_{ij} = 1}} k_{A_j \to A_i + A_j}, \\
&M_{ii} + \sum_{\substack{1 \le j \le N \\ j \neq i \\ G_{ij} = 1}} k_{A_i + A_j \to A_j} \bigg).
\end{aligned}
\end{equation}
Then we may solve \ref{eq:jacobianconstraint1} for $k_{\emptyset \to A_i}$ and \ref{eq:jacobianconstraint2} for $k_{A_i \to \emptyset}$ and obtain non-negative answers. This demonstrates that arbitrary Jacobian matrices can arise from reaction network ensembles that allow for the possibility of at least those reactions in \ref{eq:arbitraryjacobianreactionnetwork}. Note that \ref{eq:crnarbitraryjacobian}, \ref{eq:jacobianconstraint1}, \ref{eq:jacobianconstraint2}, and \ref{eq:jacobianconstraint3} are linear in the parameter values. Therefore, any probability distribution on the elements of the Jacobian can be obtained from a probability distribution on the parameter values.

\section{Hierarchy and Total ordering}\label{sec:totalordering}

A directed graph $G=(V,E)$ is a set $V$ of nodes and a set $E$ of ordered pairs of nodes \cite{Cormen2009}. For example, if $V = \{1,2,3\}$ and $E = \{(1,1),(2,2),(3,3),(1,2),(1,3),(2,3)\}$ then $G=(V,E)$ is the graph depicted in \reffigscc{} top where the labels $1$, $2$, and $3$ have been respectively assigned to the nodes vertically from top to bottom.

 We refer to the most hierarchical network architecture as the directed graph associated to a total ordering on the set of system components corresponding to the set of nodes, $V$, of the graph \cite{Cormen2009}. In general, a totally ordered set is a pair $(S,R)$ consisting of a set $S$ together with a total order relation $R$ on it. An example of a total ordering is the less than or equal to relation, $R \equiv \leq$, on the subset of natural numbers $S \equiv \{1,2,3\}$ given by $R \equiv \{1 \leq 1, 2 \leq 2, 3 \leq 3, 1 \leq 2, 1 \leq 3, 2 \leq 3\}$. The graph associated to this relation is equivalent to the graph shown in \reffigscc{} top and described algebraically in the preceding paragraph. More precisely, the conditions on $R$ for arbitrary elements $x,\,y,\,z \in S$ necessary for $(S,R)$ to be a totally ordered set are
\begin{enumerate}
\item If $x R y$ and $y R x$ then $x=y$ (antisymmetry)
\item If $x R y$ and $y R z$ then $x R z$ (transitivity)
\item $x R y$ or $y R x$ (totality)
\end{enumerate}
The totality condition implies $x R x$ (reflexivity) corresponding to the fact that the directed graph associated to the total ordering has, for each node, an edge whose source and target are the same node.

Corresponding to the SCC decomposition of $G$ we can construct a directed acyclic graph $\hier (G)$ or the condensed graph \cite{Corominas-Murtra2013}.  Each node of $\hier (G)$ corresponds to a strongly connected component of $G$. There is an edge from the node corresponding to component $C$ to the node corresponding to component $C'$ if and only if there exists a link from some vertex in $C$ to some vertex in $C'$ in $G$.  Because of the maximality of strongly connected components, $\hier (G)$ is acyclic.

The relationship between $G$ and $\hier(G)$ for all $G$ with a given number of vertices suggests a heuristic method of quantifying the degree of hierarchy of a given graph and thus of the system structure it represents. The most hierarchical system is considered to be the graph corresponding to the total ordering, which for three nodes is given in \reffigscc{} (top panel). This graph maximizes the number of links between strongly connected components, which also implies maximizing the number of strongly connected components. The graph edit distance (ED) on a fixed number of vertices from one graph to another is defined as the minimum number of modifications of the first graph in order to transform it into the second \cite{Axenovich2011}. This distance between any given graph and the total ordering thus quantitatively represents how far a graph is from being maximally hierarchical. In this work we take $max(ED) - ED$ to be the definition of hierarchy, where $max(ED)$ is the maximum edit distance for all graphs with a given number of nodes.

\pagebreak
\onecolumngrid
\pagebreak

\pagebreak

\pagebreak

\pagebreak
\FloatBarrier

\pagebreak
\FloatBarrier

\newcommand{\specialcell}[2][c]{  \begin{tabular}[#1]{@{}c@{}}#2\end{tabular}}

\begin{table*}[h]
\begin{center}
\begin{tabular}{ c || c | c | c }
\hline
matrix & connectivity & \specialcell{robustness} & \specialcell{probability\\of stability}\\
\hline
  $\begin{pmatrix}
a & b \\
d & c
\end{pmatrix}$ & 4 & 0.62 & 0.25 \\
  $\begin{pmatrix}
a & b \\
d & 0
\end{pmatrix}$, $\begin{pmatrix}
0 & b \\
d & c
\end{pmatrix}$ & 3 & 0.5 & 0.25 \\
  $\begin{pmatrix}
a & 0 \\
d & c
\end{pmatrix}$, $\begin{pmatrix}
a & b \\
0 & c
\end{pmatrix}$ & 3 & 0.67 & 0.25 \\
$\begin{pmatrix}
a & 0 \\
0 & c
\end{pmatrix}$ & 2 & 0.5 & 0.25 \\
\end{tabular}
\end{center}
\caption{{\bf Probability of stability under resampling and \emph{a priori} stability for two variable systems derived analytically}. All matrices not listed have $0$ probability of stability.}\label{tab:structstabmat}
\end{table*}

\pagebreak
\begin{longtable*}{ c | c || c | c | c | c | c }
\hline
matrix & \specialcell{orbit\\size} & connectivity & \specialcell{edit\\distance} & \specialcell{cycle\\number} & \specialcell{robustness} & \specialcell{probability\\of stability}\\
\hline
$\begin{pmatrix}
1 & 0 & 0\\
0 & 1 & 0\\
0 & 0 & 1\\
\end{pmatrix}$ & 1 & 3 & 3 & 0 & 0.499 & 0.126\\
$\begin{pmatrix}
0 & 0 & 1\\
0 & 1 & 0\\
1 & 0 & 1\\
\end{pmatrix}$ & 6 & 4 & 4 & 1 & 0.505 & 0.121\\
$\begin{pmatrix}
1 & 0 & 0\\
0 & 1 & 1\\
0 & 0 & 1\\
\end{pmatrix}$ & 6 & 4 & 2 & 0 & 0.622 & 0.127\\
$\begin{pmatrix}
0 & 0 & 1\\
0 & 1 & 1\\
1 & 0 & 1\\
\end{pmatrix}$ & 12 & 5 & 3 & 1 & 0.595 & 0.121\\
$\begin{pmatrix}
0 & 0 & 1\\
0 & 1 & 1\\
1 & 1 & 0\\
\end{pmatrix}$ & 6 & 5 & 5 & 2 & 0.494 & 0.128\\
$\begin{pmatrix}
0 & 1 & 1\\
0 & 0 & 1\\
1 & 0 & 1\\
\end{pmatrix}$ & 12 & 5 & 3 & 2 & 0.41 & 0.061\\
$\begin{pmatrix}
0 & 1 & 0\\
0 & 1 & 1\\
1 & 0 & 1\\
\end{pmatrix}$ & 6 & 5 & 3 & 1 & 0.43 & 0.078\\
$\begin{pmatrix}
0 & 1 & 1\\
0 & 1 & 0\\
1 & 0 & 1\\
\end{pmatrix}$ & 12 & 5 & 3 & 1 & 0.605 & 0.12\\
$\begin{pmatrix}
0 & 1 & 1\\
0 & 1 & 1\\
1 & 0 & 0\\
\end{pmatrix}$ & 6 & 5 & 3 & 2 & 0.405 & 0.06\\
$\begin{pmatrix}
1 & 0 & 1\\
0 & 1 & 1\\
0 & 0 & 1\\
\end{pmatrix}$ & 6 & 5 & 1 & 0 & 0.698 & 0.122\\
$\begin{pmatrix}
1 & 0 & 0\\
0 & 1 & 1\\
0 & 1 & 1\\
\end{pmatrix}$ & 3 & 5 & 3 & 1 & 0.587 & 0.128\\
$\begin{pmatrix}
1 & 0 & 1\\
0 & 1 & 0\\
0 & 1 & 1\\
\end{pmatrix}$ & 6 & 5 & 1 & 0 & 0.707 & 0.127\\
$\begin{pmatrix}
0 & 0 & 1\\
0 & 1 & 1\\
1 & 1 & 1\\
\end{pmatrix}$ & 6 & 6 & 4 & 2 & 0.578 & 0.121\\
$\begin{pmatrix}
0 & 1 & 1\\
0 & 0 & 1\\
1 & 1 & 1\\
\end{pmatrix}$ & 6 & 6 & 4 & 3 & 0.487 & 0.081\\
$\begin{pmatrix}
0 & 1 & 1\\
0 & 1 & 1\\
1 & 0 & 1\\
\end{pmatrix}$ & 12 & 6 & 2 & 2 & 0.543 & 0.098\\
$\begin{pmatrix}
0 & 1 & 0\\
0 & 1 & 1\\
1 & 1 & 1\\
\end{pmatrix}$ & 6 & 6 & 2 & 2 & 0.501 & 0.088\\
$\begin{pmatrix}
0 & 1 & 1\\
0 & 1 & 0\\
1 & 1 & 1\\
\end{pmatrix}$ & 12 & 6 & 2 & 1 & 0.662 & 0.123\\
$\begin{pmatrix}
0 & 1 & 1\\
0 & 1 & 1\\
1 & 1 & 0\\
\end{pmatrix}$ & 12 & 6 & 4 & 3 & 0.467 & 0.079\\
$\begin{pmatrix}
0 & 1 & 1\\
1 & 1 & 0\\
1 & 0 & 1\\
\end{pmatrix}$ & 3 & 6 & 4 & 2 & 0.583 & 0.13\\
$\begin{pmatrix}
1 & 0 & 1\\
0 & 1 & 1\\
0 & 1 & 1\\
\end{pmatrix}$ & 12 & 6 & 2 & 1 & 0.659 & 0.124\\
$\begin{pmatrix}
1 & 1 & 1\\
0 & 1 & 1\\
0 & 0 & 1\\
\end{pmatrix}$ & 6 & 6 & 0 & 0 & 0.751 & 0.124\\
$\begin{pmatrix}
1 & 1 & 0\\
0 & 1 & 1\\
1 & 0 & 1\\
\end{pmatrix}$ & 2 & 6 & 2 & 1 & 0.604 & 0.097\\
$\begin{pmatrix}
0 & 1 & 1\\
0 & 1 & 1\\
1 & 1 & 1\\
\end{pmatrix}$ & 12 & 7 & 3 & 3 & 0.564 & 0.103\\
$\begin{pmatrix}
0 & 1 & 1\\
1 & 0 & 1\\
1 & 1 & 1\\
\end{pmatrix}$ & 3 & 7 & 5 & 5 & 0.475 & 0.068\\
$\begin{pmatrix}
0 & 1 & 1\\
1 & 1 & 1\\
1 & 0 & 1\\
\end{pmatrix}$ & 6 & 7 & 3 & 3 & 0.591 & 0.108\\
$\begin{pmatrix}
1 & 1 & 1\\
0 & 1 & 1\\
0 & 1 & 1\\
\end{pmatrix}$ & 6 & 7 & 1 & 1 & 0.717 & 0.119\\
$\begin{pmatrix}
1 & 0 & 1\\
0 & 1 & 1\\
1 & 1 & 1\\
\end{pmatrix}$ & 3 & 7 & 3 & 2 & 0.648 & 0.122\\
$\begin{pmatrix}
1 & 1 & 1\\
0 & 1 & 1\\
1 & 0 & 1\\
\end{pmatrix}$ & 6 & 7 & 1 & 2 & 0.627 & 0.105\\
$\begin{pmatrix}
0 & 1 & 1\\
1 & 1 & 1\\
1 & 1 & 1\\
\end{pmatrix}$ & 3 & 8 & 4 & 5 & 0.577 & 0.093\\
$\begin{pmatrix}
1 & 1 & 1\\
0 & 1 & 1\\
1 & 1 & 1\\
\end{pmatrix}$ & 6 & 8 & 2 & 3 & 0.639 & 0.109\\
$\begin{pmatrix}
1 & 1 & 1\\
1 & 1 & 1\\
1 & 1 & 1\\
\end{pmatrix}$ & 1 & 9 & 3 & 5 & 0.638 & 0.106\\
\caption{{\bf Robustness and stability for three variable systems estimated via Monte Carlo sampling.} All matrices not listed have $0$ probability of stability.}\label{tab:structstabmat3}
\end{longtable*}

\end{document}